\documentclass[onecolumn,draftcls]{IEEEtran}

\usepackage{cite}
\usepackage{amsfonts}

\usepackage{graphicx}

\graphicspath{{eps/}} \DeclareGraphicsExtensions{.eps}
\usepackage[noblocks]{authblk}
\usepackage[cmex10]{amsmath}
\usepackage{amssymb}
\usepackage{url}
\ifCLASSINFOpdf

\else

\fi

\begin{document}

\title{Improving Energy Efficiency Through Multimode Transmission in the Downlink MIMO
Systems}
\author[1]{Jie Xu}
\author[1,*]{Ling Qiu}
\author[2]{Chengwen Yu}
\affil[1]{Personal Communication Network \& Spread Spectrum
Laboratory (PCN\&SS), University of Science and Technology of China
(USTC), Hefei, Anhui, 230027, China} \affil[2]{ Wireless research,
Huawei Technologies Co. Ltd., Shanghai, China}
\affil[*]{Corresponding author}

\markboth{EURASIP Journal on Wireless Communications and Networking (2nd Revision)}{Jie Xu \lowercase{\textit{et al}}.: Energy Efficient
Multimode Transmission in the Downlink MIMO Systems}

\maketitle

\begin{abstract}
Adaptively adjusting system parameters including bandwidth, transmit power and mode  to maximize the "Bits per-Joule" energy efficiency (BPJ-EE)
in the downlink MIMO systems with imperfect channel state information at the transmitter (CSIT) is considered in this paper. By mode we refer to
choice of transmission schemes i.e. singular value decomposition (SVD) or block diagonalization (BD), active transmit/receive antenna number and
active user number. We derive optimal bandwidth and transmit power for each dedicated mode at first, in which accurate capacity estimation
strategies are proposed to cope with the imperfect CSIT caused capacity prediction problem. Then, an ergodic capacity based mode switching
strategy is proposed to further improve the BPJ-EE, which provides insights on the preferred mode under given scenarios. Mode switching
compromises different power parts, exploits the tradeoff between the multiplexing gain and the imperfect CSIT caused inter-user interference,
improves the BPJ-EE significantly.
\end{abstract}

\begin{keywords}
"Bits per-Joule" energy efficiency (BPJ-EE), downlink MIMO systems,
singular value decomposition (SVD), block diagonalization (BD),
imperfect CSIT.
\end{keywords}

\section{Introduction}

\IEEEPARstart{E}{nergy} efficiency is becoming increasingly important for the future radio access networks due to the climate change and the
operator's increasing operational cost. As base stations (BSs) take the main parts of the energy consumption \cite{Fettweis1,Blume1}, improving
the energy efficiency of BS is significant. Additionally, multiple-input multiple-output (MIMO) has become the key technology in the next
generation broadband wireless networks such as WiMAX and 3GPP-LTE. Therefore, we will focus on the maximizing energy efficiency problem in the
downlink MIMO systems in this paper.

Previous works mainly focused on maximizing energy efficiency in the single-input single output (SISO) systems
\cite{huawei_vtc,Kim1,Miao1,Miao2,Miao3} and point to point single user (SU) MIMO systems \cite{Cui,Kim2,HSKim}. In the uplink TDMA SISO
channels, the optimal transmission rate was derived for energy saving in the non-real time sessions \cite{Kim1}. G. W. Miao \emph{et al.}
considered the optimal rate and resource allocation problem in OFDMA SISO channels \cite{Miao1,Miao2,Miao3}. The basic idea of
\cite{Kim1,Miao1,Miao2,Miao3} is finding an optimal transmission rate to compromise the power amplifier (PA) power which is proportional to the
transmit power and the circuit power which is independent of the transmit power. S. Zhang \emph{et al.} extended the energy efficiency problem
to a bandwidth variable system \cite{huawei_vtc} and the bandwidth-power-energy efficiency relations were investigated. As the MIMO systems can
improve the data rates compared with SISO/SIMO, the transmit power can be reduced under the same rate. Meanwhile, MIMO systems consume higher
circuit power than SISO/SIMO due to the multiplicity of associated circuits such as mixers, synthesizers, digital-to-analog converters (DAC),
filters, etc. \cite{Cui} is the pioneering work in this area comparing the energy efficiency of Alamouti MIMO systems with 2 antennas and SIMO
systems in the sensor networks. H. Kim \emph{et al.} presented the energy efficient mode switching between SIMO and 2 antenna MIMO systems
\cite{Kim2}. A more general link adaptation strategy was proposed in \cite{HSKim} and the system parameters including number of data streams,
number of transmit/receive antennas, use of spatial multiplexing or space time block coding (STBC), bandwidth etc. were controlled to maximize
the energy efficiency. However, to the best of our knowledge, there are few literatures considering energy efficiency of the downlink multiuser
(MU) MIMO systems.

The number of transmit antennas at BS is always larger than the number of receive antennas at the mobile station (MS) side because of the MS's
size limitation. MU-MIMO systems can provide higher data rates than SU-MIMO by transmitting to multiple MSs simultaneously over the same
spectrum. Previous literatures mainly focused on maximizing the spectral efficiency of MU-MIMO systems, some examples of which are
\cite{Spencer,Shen,Rchen1,Rchen2,Zhang1,Zhang2,Zhang3,Xu2}. Although not capacity achieving, block diagonalization (BD) is a popular linear
precoding scheme in the MU-MIMO systems \cite{Spencer,Shen,Rchen1,Rchen2}. Performing precoding requires the channel state information at the
transmitter(CSIT) and the accuracy of CSIT impacts the performance significantly. The imperfect CSIT will cause inter-user interference and the
spectral efficiency will decrease seriously. In order to compromise the spatial multiplexing gain and the inter-user interference, spectral
efficient mode switching between SU-MIMO and MU-MIMO was presented in \cite{Zhang1,Zhang2,Zhang3,Xu2}.

Maximizing the "Bits per-Joule" energy efficiency (BPJ-EE) in the downlink MIMO systems with imperfect CSIT is addressed in this paper. A three
part power consumption model is considered. By power conversion (PC) power we refer to power consumption proportional to the transmit power,
which captures the effect of PA, feeder loss, and extra loss in transmission related cooling. By static power we refer to the power consumption
which is assumed to be constant irrespective of the transmit power, number of transmit antennas and bandwidth. By dynamic power we refer to the
power consumption including the circuit power, signal processing power, etc. and it is assumed to be irrespective of the transmit power but
dependent of the number of transmit antennas and bandwidth. We divide the dynamic power into three parts. The first part "Dyn-I" is proportional
to the transmit antenna number only, which can be viewed as the circuit power. The second part "Dyn-II" is proportional to the bandwidth only,
and the third part "Dyn-III" is proportional to the multiplication of the bandwidth and transmit antenna number. "Dyn-II" and "Dyn-III" can be
viewed as the signal processing power etc.. Interestingly, there are two main tradeoffs here. For one thing, more transmit antennas would
increase the spatial multiplexing and diversity gain which leads to transmit power saving, while more transmit antennas would increase "Dyn-I"
and "Dyn-III" which leads to dynamic power wasting. For another, multiplexing more active users with higher multiplexing gain would increase the
inter-user interference, in which the multiplexing gain makes transmit power saving but inter-user interference induces transmit power wasting.
In order to maximize BPJ-EE, the tradeoff among PC power, static power and dynamic power needs to be resolved and the tradeoff between the
multiplexing gain and imperfect CSIT caused inter-user interference also needs to be carefully studied. The optimal adaptation which adaptively
adjusts system parameters such as bandwidth, transmit power, use of singular value decomposition (SVD) or BD, number of active transmit/receive
antennas, number of active users is considered in this paper to meet the challenge.

{{{\bf{Contributions.}}}} By mode we refer to choice of transmission schemes i.e. SVD or BD, active transmit/receive antenna number and active
user number. For each dedicated mode, we prove that the BPJ-EE is monotonically increasing as a function of bandwidth under the optimal transmit
power without maximum power constraint. Meanwhile, we derive the unique globally optimal transmit power with a constant bandwidth. Therefore,
the optimal bandwidth is chosen to use the whole available bandwidth and the optimal transmit power can be correspondingly obtained. However,
due to imperfect CSIT, it is emphasized that the capacity prediction is a big challenge during the above derivation. To cope with this problem,
a capacity estimation mechanism is presented and accurate capacity estimation strategies are proposed.

The derivation of the optimal transmit power and bandwidth reveals the relationship between the BPJ-EE and the mode. Applying the derived
optimal transmit power and bandwidth, mode switching is addressed then to choose the optimal mode. An ergodic capacity based mode switching
algorithm is proposed. We derive the accurate close-form capacity approximation for each mode under imperfect CSIT at first and calculate the
optimal BPJ-EE of each mode based on the approximation. Then, the preferred mode can be decided after comparison. The proposed mode switching
scheme provides guidance on the preferred mode under given scenarios and can be applied offline. Simulation results show that the mode switching
improves the BPJ-EE significantly and it is promising for the energy efficient transmission.

The rest of the paper is organized as follows. Section {{\MakeUppercase{\romannumeral 2}}} introduces the system model, power model and two
transmission schemes and then section {{\MakeUppercase{\romannumeral 3}}} gives the problem definition. Optimal bandwidth, transmit power
derivation for each dedicated mode and capacity estimation under imperfect CSIT are presented in section {{\MakeUppercase{\romannumeral 4}}}.
The ergodic capacity based mode switching is proposed in section {{\MakeUppercase{\romannumeral 5}}}. The simulation results are shown in
section {{\MakeUppercase{\romannumeral 6}}} and finally, section {{\MakeUppercase{\romannumeral 7}}} concludes this paper.

Regarding the notation, bold face letters refer to vectors (lower
case) or matrices (upper case). Notation ${\mathbb{E}}(\bf{A})$ and
${\rm{Tr}}(\bf{A})$ denote the expectation and trace operation of
matrix $\bf{A}$, respectively. The superscript $H$ and $T$ represent
the conjugate transpose and transpose operation, respectively.

\section{Preliminaries}

\subsection{System model}

The downlink MIMO systems consist of a single BS with $M$ antennas and $K$ users each with $N$ antennas. $M \geq K \times N$ is assumed. We
assume that the channel matrix from the BS to the $k$th user at time $n$ is ${\bf{H}}_{k}[n] \in {\mathbb{C}}^{N \times M},k=1,\ldots,K$, which
can be denoted as
\begin{equation} \label{eq1}
\begin{array}{l}
{\bf{H}}_k[n] = \zeta_k {\hat{\bf{H}}}_k[n] = \Phi d_k ^ {-\lambda}
\Psi {\hat{\bf{H}}}_k[n].
\end{array}
\end{equation}
$\zeta_k  = \Phi d_k ^ {-\lambda} \Psi$ is the large scale fading including pathloss and shadowing fading, in which $d_k$, $\lambda$ denote the
distance from the BS to the user $k$ and the path loss exponent, respectively. The random variable $\Psi$ accounts for the shadowing process.
The term $\Phi$ denotes the pathloss parameter to further adapt the model which accounts for the BS and MS antenna heights, carrier frequency,
propagation conditions and reference distance. ${\hat{\bf{H}}}_k[n]$ denotes the small scale fading channel. We assume that the channel
experiences flat fading and ${\hat{\bf{H}}}_k[n]$ is well modeled as a spatially white Gaussian channel, with each entry ${\mathcal {C}\mathcal
{N}}(0,1)$.

For the $k$th user, the received signal can be denoted as
\begin{equation} \label{eq2}
\begin{array}{l}
{\bf{y}}_k[n] = {\bf{H}}_k[n] {\bf{x}}[n] + {\bf{n}}_k[n],
\end{array}
\end{equation}
in which ${\bf{x}}[n] \in {\mathbb{C}}^{M \times 1}$ is the BS's transmitted signal, ${\bf{n}}_k[n]$ is the Gaussian noise vector with entries
distributed according to ${\mathcal {C}\mathcal {N}}(0,N_0W)$, where $N_0$ is the noise power density and $W$ is the carrier bandwidth. The
design of ${\bf{x}}[n]$ depends on the transmission schemes which would be introduced in subsection \ref{Trans_schemes}.

As one objective of this paper is to study the impact of imperfect
CSIT, we will assume perfect channel state information at the
receive (CSIR) and imperfect CSIT here. CSIT is always get through
feedback from the MSs in the FDD systems and through uplink channel
estimation based on uplink-downlink reciprocity in the TDD systems,
so the main sources of CSIT imperfection come from channel
estimation error, delay and feedback error
\cite{Zhang1,Zhang2,Zhang3}. Only  the delayed CSIT imperfection is
considered in this paper, but note that the delayed CSIT model can
be simply extended to other imperfect CSIT case such as estimation
error and analog feedback \cite{Zhang1,Zhang2}. The channels will
stay constant for a symbol duration and change from symbol to symbol
according to a stationary correlation model. Assume that there is
$D$ symbols delay between the estimated channel and the downlink
channel. The current channel ${\bf{H}}_k[n] = \zeta_k
{\hat{\bf{H}}}_k[n]$  and its delayed version ${\bf{H}}_k[n-D] =
\zeta_k {\hat{\bf{H}}}_k[n-D]$ are jointly Gaussian with zero mean
and are related in the following manner \cite{Zhang2}.
\begin{equation} \label{eq3}
\begin{array}{l}
{\hat{\bf{H}}}_k[n] = \rho_k{\hat{\bf{H}}}_k[n-D] + {\hat{\bf{E}}}_k[n],
\end{array}
\end{equation}
where $\rho_k$ denotes the correlation coefficient of each user, ${\hat{\bf{E}}}_k[n]$ is the channel error matrix, with i.i.d. entries
${\mathcal {C}\mathcal {N}}(0,{\epsilon}_{e,k}^2)$ and it is uncorrelated with $\hat{{\bf{H}}}_k[n-D]$. Meanwhile, we denote ${{\bf{E}}}_k[n] =
\zeta_k{\hat{\bf{E}}}_k[n]$. The amount of delay is $\tau = D T_s$, where $T_s$ is the symbol duration. And $\rho_k = J_0(2\pi f_{d,k}\tau)$
with Doppler spread $f_{d,k}$, where $J_0(\cdot)$ is the zeroth order Bessel function of the first kind, and $\epsilon^2_{e,k} = 1- \rho_k^2$
\cite{Zhang2}. Therefore, both $\rho_k$ and $\epsilon_{e,k}$ are determined by the normalized Doppler frequency $f_{d,k}\tau$.

\subsection{Power model}

Apart from PA power and the circuit power, the signal processing, power supply and air condition power should also be taken into account at the
BS \cite{Arnold}. Before introduction, assume the number of active transmit antennas is $M_{\rm{a}}$ and the total transmit power is
$P_{\rm{t}}$. Motivated by the power model in \cite{Arnold,huawei_vtc,HSKim}, the three part power model is introduced as follows. The total
power consumption at BS is divided into three parts. The first part is the PC power
\begin{equation} \label{eq4_0}
\begin{array}{l}P_{\rm{PC}} = \frac{P_{\rm{t}}}{\eta},
\end{array}
\end{equation}
in which $\eta$ is the power conversion efficiency, accounting for the PA efficiency, feeder loss, and extra loss in transmission related
cooling. Although the total transmit power should be varied as $M_{\rm{a}}$ and $W$ changes, we study the total transmit power as a whole and
the PC power includes all the total transmit power. The effect of $M_{\rm{a}}$ and $W$ on the transmit power independent power is expressed by
the second part: the dynamic power $P_{\rm{Dyn}}$. $P_{\rm{Dyn}}$ captures the effect of signal processing, circuit power, etc., which is
dependent of $M_{\rm{a}}$ and $W$ but independent of $P_{\rm{t}}$. $P_{\rm{Dyn}}$ is separated into three classes. The first class "Dyn-I"
$P_{{\rm{Dyn-I}}}$ is proportional to the transmit antenna number only, which can be viewed as the circuit power of the RF. The second part
"Dyn-II" $P_{{\rm{Dyn-II}}}$ is proportional to the bandwidth only, and the third part "Dyn-III" $P_{{\rm{Dyn-III}}}$ is proportional to the
multiplication of the bandwidth and transmit antenna number. $P_{{\rm{Dyn-II}}}$ and $P_{{\rm{Dyn-III}}}$ can be viewed as the signal processing
related power. Thus, the dynamic power can be denoted as follows.
\begin{equation} \label{eq4_1}
\begin{array}{l}
P_{\rm{Dyn}} = P_{{\rm{Dyn-I}}} + P_{{\rm{Dyn-II}}} +
P_{{\rm{Dyn-III}}},\\
P_{{\rm{Dyn-I}}} = M_{\rm{a}} P_{\rm{cir}},\\
P_{{\rm{Dyn-II}}} = p_{\rm{ac,bw}} W, \\
P_{{\rm{Dyn-III}}} = M_{\rm{a}} p_{\rm{sp,bw}}   W,
\end{array}
\end{equation}
The third part is the static power $P_{\rm{Sta}}$ which is independent of $P_{\rm{t}}$, $M_{\rm{a}}$ and $W$, including the power consumption of
cooling systems, power supply and so on. Combining the three parts, we have the total power consumption as follows.
\begin{equation} \label{eq4}
\begin{array}{l}
P_{\rm{total}} = {P_{\rm{PC}}} + P_{\rm{Dyn}} + P_{\rm{Sta}}.
\end{array}
\end{equation}
Although the above power model is simple and abstract, it captures the effect of the key parameters such as $P_{\rm{t}}$, $M_{\rm{a}}$ and $W$
and coincides with the previous literatures \cite{Arnold,huawei_vtc,HSKim}. Measuring the accurate power model for a dedicated BS is very
important for the research of energy efficiency and the measuring may need careful field test, however, it is out of the scope here.

Note that here we omit the power consumption at the user side, as the users' power consumption is negligible compared with the power consumption
of BS. Although any BS power saving design should consider the impact to the users' power consumption, it is beyond the scope of this paper.

\subsection{Transmission Schemes}\label{Trans_schemes}

SU-MIMO with SVD and MU-MIMO with BD are considered in this paper as the transmission schemes. We will introduce them in this subsection.

\subsubsection{SU-MIMO with SVD}
Before discussion, we assume that $M_{{\rm{a}}}$ transmit antennas are active in the SU-MIMO. As more active receive antennas result in transmit
power saving due to higher spatial multiplexing and diversity gain, $N$ antennas should be all active at the MS side{\footnote{Here more receive
antenna at MS will cause higher MS power consumption. However, note that the power consumption of MS is omitted.}}. The number of data streams
is limited by the minimum number of transmit and receive antennas, which is denoted as $N_{\rm{s}} = \min(M_{\rm{a}},N)$.

In the SU-MIMO mode, SVD with equal power allocation is applied. Although SVD with waterfilling is the capacity optimal scheme \cite{Telatar},
considering equal power allocation here helps the comparison between SU-MIMO and MU-MIMO fairly \cite{Zhang2}. The SVD of ${\bf{H}}[n]$ is
denoted as
\begin{equation} \label{eq8}
\begin{array}{l}
{\bf{H}}[n] = {\bf{U}}[n]{\bf{\Lambda}}[n]{\bf{V}}[n]^H,
\end{array}
\end{equation}
in which ${\bf{\Lambda}}[n]$ is a diagonal matrix, ${\bf{U}}[n]$ and
${\bf{V}}[n]$ are unitary. The precoding matrix is designed as
${\bf{V}}[n]$ at the transmitter in the perfect CSIT scenario.
However, when only the delayed CSIT is available at the BS, the
precoding matrix is based on the delayed version which should be
${\bf{V}}[n-D]$. After the MS preforms MIMO detection, the
achievable capacity can be denoted as
\begin{equation} \label{eq9}
\begin{array}{l}
R_{\rm{s}}({M_{\rm{a}}},{P_{\rm{t}}},W) =W\sum \limits
_{i=1}^{{N_{\rm{s}}}}\log\left( 1 +
\frac{P_{\rm{t}}}{{N_{\rm{s}}}{N_0W}} \lambda^2_i \right),
\end{array}
\end{equation}
where $\lambda_i$ is the $i$th singular value of
${\bf{H}}[n]{\bf{V}}[n-D]$.

\subsubsection{MU-MIMO with BD}

We assume that $K_{\rm{a}}$ users each with $N_{{\rm{a}},i}, i=1,\ldots,K_{\rm{a}}$ antennas are active at the same time. Denote the total
receive antenna number as $N_{\rm{a}} = \sum \limits _{i=1}^{K_{\rm{a}}}N_{{\rm{a}},i}$. As linear precoding is preformed, we have that
$M_{\rm{a}} \geq N_{\rm{a}}$\cite{Spencer}, and then the number of data streams is $N_{\rm{s}} = N_{\rm{a}}$. The BD precoding scheme with equal
power allocation is applied in the MU-MIMO mode. Assume that the precoding matrix for the $k$th user is ${\bf{T}}_k[n]$ and the desired data for
the $k$th user is ${\bf{s}}_k[n]$, then ${\bf{x}}[n] = \sum\limits_{i=1}^{K_{\rm{a}}}{\bf{T}}_i[n]{\bf{s}}_i[n]$. The transmission model is
\begin{equation} \label{eq11}
\begin{array}{l}
{\bf{y}}_k[n] = {\bf{H}}_k[n]
\sum\limits_{i=1}^{K_{\rm{a}}}{\bf{T}}_i[n]{\bf{s}}_i[n] +
{\bf{n}}_k[n].
\end{array}
\end{equation}

In the perfect CSIT case, the precoding matrix is based on ${\bf{H}}_k[n] \sum\limits_{i=1,i \neq k}^{K_{\rm{a}}}{\bf{T}}_i[n] = {\bf{0}}$. The
detail of the design can be found in \cite{Spencer}. Define the effective channel as ${\bf{H}}_{{\rm{eff}},k}[n] =
{\bf{H}}_{k}[n]{\bf{T}}_k[n]$. Then the capacity can be denoted as
\begin{equation} \label{eq12}
\begin{array}{l}
R_{\rm{b}}^P
({M_{\rm{a}}},{K_{\rm{a}}},{N_{\rm{a,1}}},\ldots,{N_{\rm{a,{K_{\rm{a}}}}}},{P_{\rm{t}}},W)
=\\W\sum\limits_{k=1}^{K_{\rm{a}}}\log\det\left( {\bf{I}} +
\frac{P_{\rm{t}}}{{N_{\rm{s}}}{N_0W}}{\bf{H}}_{{\rm{eff}},k}[n]
{\bf{H}}_{{\rm{eff}},k}^H [n] \right).
\end{array}
\end{equation}

In the delayed CSIT case, the precoding matrix design is based on the delayed version, i.e. ${\bf{H}}_k[n-D] \sum\limits_{i=1,i \neq
k}^{K_{\rm{a}}}{\bf{T}}_i^{(D)}[n] = {\bf{0}}$. Then define the effective channel in the delayed CSIT case as $\hat{\bf{H}}_{{\rm{eff}},k}[n] =
{\bf{H}}_{k}[n]{\bf{T}}^{(D)}_k[n]$. The capacity can be denoted as \cite{Zhang2}
\begin{equation} \label{eq13}
\begin{array}{l}
R_{\rm{b}}^D
({M_{\rm{a}}},{K_{\rm{a}}},{N_{\rm{a,1}}},\ldots,{N_{\rm{a,{K_{\rm{a}}}}}},{P_{\rm{t}}},W)
=\\W\sum\limits_{k=1}^{K_{\rm{a}}}\log\det\left( {\bf{I}} +
\frac{P_{t}}{{N_{\rm{s}}}}\hat{\bf{H}}_{{\rm{eff}},k}[n]\hat{\bf{H}}_{{\rm{eff}},k}^H
[n]{\bf{R}}_k^{-1} [n] \right),
\end{array}
\end{equation}
in which
\begin{equation} \label{eq14}
\begin{array}{l}
{\bf{R}}_k[n] = \frac{P_{\rm{t}}}{N{\rm{s}}}{\bf{E}}_k[n]\left[ \sum
\limits_{i\neq k} {\bf{T}}_i^{(D)}[n] {\bf{T}}_i^{(D)H} [n] \right]
{\bf{E}}_k^H [n] + N_0W{\bf{I}}
\end{array}
\end{equation}
is the inter-user interference plus noise part.

\section{Problem Definition}
The objective of this paper is to maximize the BPJ-EE in the downlink MIMO systems. The BPJ-EE is defined as the achievable capacity divided by
the total power consumption, which is also the transmitted bits per unit energy (Bits/Joule). Denote the BPJ-EE as $\xi$ and then the
optimization problem can be denoted as
\begin{equation} \label{eq5}
\begin{array}{l}
\mathop {\max }{\xi} =  \frac{R_{\rm{m}}
({M_{\rm{a}}},{K_{\rm{a}}},{N_{\rm{a,1}}},\ldots,{N_{{{\rm{a}},{K_{\rm{a}}}}}},{P_{\rm{t}}},W)}{{{P_{{\rm{total}}}}}} \\
{\rm{s.t.}}{\;} P_{\rm{TX}}\geq 0, \\
{\quad}{\,}{\,\,} 0\leq W\leq W_{\rm{max}}.
\end{array}
\end{equation}
According to the above problem, bandwidth limitation is considered. In order to make the transmission most energy efficient, we should
adaptively adjust the following system parameters: transmission scheme ${\rm{m}}\in \{{\rm{s}},{\rm{b}}\}$, i.e. use of SVD or BD, number of
active transmit antennas ${M_{\rm{a}}}$,  number of active users ${K_{\rm{a}}}$, number of receive antennas ${N_{{\rm{a}},i}}, i =1,\ldots,
{K_{\rm{a}}}$, transmit power ${P_{\rm{t}}}$ and bandwidth $W$.

The optimization of problem (\ref{eq5}) is divided into two steps. At first, determine the optimal ${P_{\rm{t}}}$ and $W$ for each dedicated
mode. After that, apply mode switching to determine the optimal mode, i.e. optimal transmission scheme ${\rm{m}}$, optimal transmit antenna
number ${M_{\rm{a}}}$, optimal user number ${K_{\rm{a}}}$ and optimal receive antenna number ${N_{{\rm{a}},i}}$, according to the derivations of
the first step. The next two sections will describe the details.

\section{Maximizing Energy Efficiency with Optimal Bandwidth and Transmit Power}

The optimal bandwidth and transmit power are derived in this section under a dedicated mode. Unless otherwise specified, the mode, i.e.
transmission scheme ${\rm{m}}$, active transmit antenna number ${M_{\rm{a}}}$, active receive antenna number ${N_{{\rm{a}},i}}, i =1,\ldots,
{K_{\rm{a}}}$ and active user number ${K_{\rm{a}}}$, is constant in this section. The following lemma is introduced at first to help the
derivation.

\newtheorem{Lemma}{Lemma}
\begin{Lemma}\label{lemma0}
For optimization problem
\begin{equation} \label{eq5_0}
\begin{array}{l}
\max \frac{f(x)}{ax+b}, \\
{\rm{s.t.}} \; x\geq 0
\end{array}
\end{equation}
in which $a>0$ and $b>0$. $f(x) \geq 0\left( x \geq 0\right)$ and $f(x)$ is strictly concave and monotonically increasing. There exists a unique
globally optimal $x^{*}$ given by
\begin{equation} \label{eq5_1}
\begin{array}{l}
x^{*}  = \frac{f(x^{*})}{f^{'}(x^{*})}- \frac{b}{a},
\end{array}
\end{equation}
where $f^{'}(x)$ is the first derivative of function $f(x)$.
\end{Lemma}

{\bf{Proof:}} See Appendix \ref{A_1}.

\subsection{Optimal Energy Efficient Bandwidth}\label{EE_bandwidth}
To illustrate the effect of bandwidth on the BPJ-EE, the following theorem is derived.

\newtheorem{Theorem}{Theorem}
\begin{Theorem}\label{Theorem0}
Under constant $P_{\rm{t}}$, there exists a unique globally optimal
$W^*$ given by
\begin{equation} \label{eq5_3}
\begin{array}{l}
W^{*}  = \frac{(P_{\rm{PC}}+P_{\rm{Sta}}+M_{\rm{a}}P_{\rm{cir}})+(M_{\rm{a}}p_{\rm{sp,bw}}
+P_{\rm{ac,bw}})R(W^{*})}{(M_{\rm{a}}p_{\rm{sp,bw}}+P_{\rm{ac,bw}})R^{'}(W^{*})}
\end{array}
\end{equation}
to maximize $\xi$, in which $R(W)$ denotes the achievable capacity with a dedicated mode. If the transmit power scales as $P_{\rm{t}} =
p_{\rm{t}} W$, $\xi$ is monotonically increasing as a function of $W$.
\end{Theorem}

{\bf{Proof:}} See Appendix \ref{A_2}.

This Theorem provides helpful insights about the system configuration. When the transmit power of BS is fixed, configuring the optimal bandwidth
helps improve the energy efficiency. Meanwhile, if the transmit power can increase proportionally as a function of bandwidth based on
$P_{\rm{t}} = p_{\rm{t}} W$, transmitting over the whole available spectrum is thus the optimal energy efficient transmission strategy. As
$P_{\rm{t}}$ can be adjusted in problem (\ref{eq5}) and no maximum transmit power constraint is considered there, choosing $W^* = W_{\rm{max}}$
as the optimal bandwidth can maximize $\xi$. Therefore, $W^* = W_{\rm{max}}$ is applied in the rest of this paper.

One may argue that the transmit power is limited by the BS's maximum
power in the real systems. In that case, $W$ and $P_{\rm{t}}$ should
be jointly optimized. We consider this problem in our another work
\cite{Jie_Xu}.

\subsection{Optimal Energy Efficient Transmit Power}\label{optimal_selection}

After determining the optimal bandwidth, we should derive the
optimal $P_{\rm{t}}^*$ under $W^*=W_{\rm{max}}$. In this case, we
denote the capacity as $R(P_{\rm{t}})$ with the dedicated mode. Then
the optimal transmit power is derived according to the following
theorem.

\begin{Theorem}\label{Theorem1}
There exists a unique globally optimal transmit power
$P_{\rm{t}}^{*}$ of the BPJ-EE optimization problem  given by
\begin{equation} \label{eq16}
\begin{array}{l}
P_{{\rm{t}}}^{*}
 = \frac{R(P_{{\rm{t}}}^{*})}{R^{'}(P_{{\rm{t}}}^{*})}-\eta
 (P_{\rm{Sta}} + P_{\rm{Dyn}}).
\end{array}
\end{equation}
\end{Theorem}

{\bf{Proof:}} See Appendix \ref{A_3}.

Therefore, the optimal bandwidth and transmit power are derived
based on Theorem \ref{Theorem0} and Theorem \ref{Theorem1}. That is
to say, the optimal bandwidth is chosen as $W^* = W_{\rm{max}}$ and
the optimal transmit power is derived according to (\ref{eq16}).

However, note that during the optimal transmit power derivation (\ref{eq16}), the BS needs to know the achievable capacity based on the CSIT
prior to the transmission. If perfect CSIT is available at BS, the capacity formula can be calculated at the BS directly according to
(\ref{eq9}) for SU-MIMO with SVD and (\ref{eq12}) for MU-MIMO with BD. But if the CSIT is imperfect, the BS need to predict the capacity then.
In order to meet the challenge, a capacity estimation mechanism with delayed version of CSIT is developed, which is the main concern of next
subsection.

\subsection{Capacity Estimation Under Imperfect CSIT}\label{delayed_CSIT}

\subsubsection{SU-MIMO}

SU-MIMO with SVD is relatively robust to the imperfect CSIT
\cite{Zhang2}, using the delayed version of CSIT directly is a
simple and direct way. Following proposition shows the capacity
estimation of SVD mode.

\newtheorem{Proposition}{Proposition}
\begin{Proposition}\label{Proposition0}
The capacity estimation of SU-MIMO with SVD is directly estimated
by:
\begin{equation} \label{eq18}
\begin{array}{l}
R_{\rm{s}}^{{{\rm{est}}}} =W\sum \limits
_{i=1}^{{N_{\rm{s}}}}\log\left( 1 +
\frac{P_{\rm{t}}}{{N_{\rm{s}}}{N_0W}} \tilde{\lambda}^2_i \right),
\end{array}
\end{equation}
where $\tilde{\lambda}_i$ is the singular value of ${\bf{H}}[n-D]$.
\end{Proposition}

Proposition \ref{Proposition0} is motivated by \cite{Zhang2}. In Proposition \ref{Proposition0}, when the receive antenna number is equal to or
larger than the transmit antenna number, the degree of freedom can be fully utilized after the receiver's detection, and then the ergodic
capacity of (\ref{eq18}) would be the same as the delayed CSIT case in (\ref{eq9}). When the receive antenna number is smaller than the transmit
antenna number, although delayed CSIT would cause degree of freedom loss and (\ref{eq18}) cannot express the loss, the simulation will show that
Proposition \ref{Proposition0} is accurate enough to obtain the optimal $\xi$ in that case.

\subsubsection{MU-MIMO}

Since the imperfect CSIT leads to interuser interference in the
MU-MIMO systems, simply using the delayed CSIT can not accurately
estimate the capacity any longer. We should take the impact of
interuser interference into account. J. Zhang et al. first
considered the performance gap between the perfect CSIT case and the
imperfect CSIT case \cite{Zhang2}, which is described as the
following lemma.

\begin{Lemma}\label{Lemma2}
The rate loss of BD with the delayed CSIT is upper bounded
by\cite{Zhang2}:
\begin{equation} \label{eq19}
\begin{array}{l}
\triangle R_{\rm{b}} = R_{\rm{b}}^P - R_{\rm{b}}^D \leq \triangle R_{\rm{b}}^{\rm{upp}}=
\\ W\sum \limits _{k=1}^{K_{\rm{a}}} {N_{{\rm{a}},k}}
\log_2\left[ \sum\limits_{i=1,i\neq
k}^{{K_{\rm{a}}}}{N_{{\rm{a}},i}}\frac{P_{\rm{t}}\zeta_k}{N_0WN_{\rm{s}}}\epsilon_{e,k}^2
+ 1\right].
\end{array}
\end{equation}
\end{Lemma}

As the BS can get the statistic variance of the channel error $\epsilon_{e,k}^2$ due to the doppler frequency estimation, the BS can obtain the
upper bound gap $\triangle R_{\rm{b}}^{\rm{upp}}$ through some simple calculation. According to Proposition \ref{Proposition0}, we can use the
delayed CSIT to estimate the capacity with perfect CSIT $R_{\rm{b}}^P$ and we denote the estimated capacity with perfect CSIT as
\begin{equation} \label{eq19_10}
\begin{array}{l}
R_{\rm{b}}^{{\rm{est}},P} =\\
W\sum\limits_{k=1}^{K_{\rm{a}}}\log\det\big( {\bf{I}} +
\frac{P_{\rm{t}}}{{N_{\rm{s}}}{N_0W}}{\bf{H}}_{{\rm{eff}},k}[n-D]
{\bf{H}}_{{\rm{eff}},k}^H [n-D] \big),
\end{array}
\end{equation}
in which ${\bf{H}}_{{\rm{eff}},k}[n-D]={\bf{H}}_k[n-D]{\bf{T}}_k[n-D]$. Combining (\ref{eq19_10}) and Lemma \ref{Lemma2}, a lower bound capacity
estimation is denoted as the perfect case capacity $R_{\rm{b}}^{{\rm{est}},P} $ minus the capacity upper bound gap $\triangle
R_{\rm{b}}^{\rm{upp}}$, which can be denoted as \cite{Xu2}
\begin{equation} \label{eq20}
\begin{array}{l}
R_{\rm{b}}^{{\rm{est-Zhang}}} = R_{\rm{b}}^{{\rm{est}},P} - \triangle R_{\rm{b}}^{\rm{upp}}.
\end{array}
\end{equation}

However, this lower bound is not tight enough, a novel lower bound estimation and a novel upper bound estimation are proposed to estimate the
capacity of MU-MIMO with BD.

\begin{Proposition}\label{Proposition1}
The lower bound of the capacity estimation of MU-MIMO with BD is
given by (\ref{eq21}), while the upper bound of the capacity
estimation of MU-MIMO with BD is given by (\ref{eq23}). The lower
bound in (\ref{eq21}) is tighter than
$R_{\rm{b}}^{{\rm{est,Zhang}}}$ in (\ref{eq20}).
\end{Proposition}

\begin{figure*}[!ht]
\begin{equation} \label{eq21}
\begin{array}{l}
R_{\rm{b}}^{{\rm{est,low}}} =
W\sum\limits_{k=1}^{K_{\rm{a}}}\log\det\big( {\bf{I}} +
\frac{P_{\rm{t}}/{N_{\rm{s}}}}{{N_0W+\sum\limits_{i=1,i\neq
k}^{{K_{\rm{a}}}}{N_{{\rm{a}},i}}\frac{P_{\rm{t}}\zeta_k}{N_{\rm{s}}}
\epsilon_{e,k}^2}}{\bf{H}}_{{\rm{eff}},k}[n-D]{\bf{H}}_{{\rm{eff}},k}^H[n-D]
\big)
\end{array}
\end{equation}
\begin{equation} \label{eq23}
\begin{array}{l}
R_{\rm{b}}^{{\rm{est,upp}}} =
W\sum\limits_{k=1}^{K_{\rm{a}}}\left\{\log\det\big( {\bf{I}} +
\frac{P_{\rm{t}}/{N_{\rm{s}}}}{{N_0W+\sum\limits_{i=1,i\neq
k}^{{K_{\rm{a}}}}{N_{{\rm{a}},i}}\frac{P_{\rm{t}}\zeta_k}{N_{\rm{s}}}
\epsilon_{e,k}^2}}{\bf{H}}_{{\rm{eff}},k}[n-D]{\bf{H}}_{{\rm{eff}},k}^H[n-D]
\big)+ ({N_{{\rm{a}},k}}/{M_{\rm{a}}}) \log_2(e)\right\}
\end{array}
\end{equation}
\end{figure*}

Proposition \ref{Proposition1} is motivated by \cite{Yoo}. It is
illustrated as follows. Rewrite the transmission mode of user $k$ of
(\ref{eq11}) as
\begin{equation} \label{eq23_1}
\begin{array}{l}
{\bf{y}}_k[n] = {\bf{H}}_k[n] {\bf{T}}_k[n]{\bf{s}}_k[n] +
{\bf{H}}_k[n] \sum\limits_{i\neq k }{\bf{T}}_i[n]{\bf{s}}_i[n] +
{\bf{n}}_k[n].
\end{array}
\end{equation}

With delayed CSIT ,denote \[{\bf{B}}_k [n] = {\bf{H}}_k[n] \sum\limits_{i\neq k }{\bf{T}}^{(D)}_i[n]{\bf{s}}_i[n]  = {\bf{E}}_k[n]
\sum\limits_{i\neq k }{\bf{T}}^{(D)}_i[n]{\bf{s}}_i[n],\] then ${\bf{A}}_k [n] = {\bf{B}}_k [n]{\bf{B}}_k ^H[n]$ and the covariance matrix of
the interference plus noise is then
\begin{equation} \label{eq23_2}
\begin{array}{l}
{\bf{R}}_k [n] = \frac{P_{\rm{t}}}{N_{\rm{s}}}{\bf{A}}_k [n] +
N_0W{\bf{I}} [n].
\end{array}
\end{equation}
The expectation of ${\bf{R}}_k [n]$ is \cite{Zhang2}
\begin{equation} \label{eq22}
\begin{array}{l}
{\mathbb{E}}\left({\bf{R}}_k [n]\right) = \sum\limits_{i=1,i\neq
k}^{{K_{\rm{a}}}}{N_{{\rm{a}},i}}\frac{P_{\rm{t}}\zeta_k}{N_{\rm{s}}}\epsilon_{e,k}^2
{\bf{I}} + N_0W{\bf{I}}
\end{array}
\end{equation}

Based on Proposition \ref{Proposition0}, we use ${\bf{H}}_{{\rm{eff}},k}[n-D]$ with the delayed CSIT to replace the
$\hat{\bf{H}}_{{\rm{eff}},k}[n]$ in (\ref{eq13}). And then the capacity expression of each user is similar with the SU-MIMO channel with
inter-stream interference. The capacity lower bound and upper bound with a point to point MIMO channel with channel estimation errors in
\cite{Yoo} is applied here. Therefore, the lower bound estimation (\ref{eq21}) and upper bound estimation (\ref{eq23}) can be verified according
to the lower and upper bounds in \cite{Yoo} and (\ref{eq22}).

We can get $R_{\rm{b}}^{{\rm{est,low}}} -
R_{\rm{b}}^{{\rm{est,Zhang}}}
> 0$ after some simple calculation, so $R_{\rm{b}}^{{\rm{est,low}}}$
is tighter than $R_{\rm{b}}^{{\rm{est,Zhang}}}$. $\Box$

According to Proposition \ref{Proposition0} and  Proposition \ref{Proposition1}, the capacity estimation for both SVD and BD can be performed.
In order to apply Proposition \ref{Proposition0} and  Proposition \ref{Proposition1} to derive the optimal bandwidth and transmit power, it is
necessary to prove that the capacity estimation (\ref{eq18}) for SU-MIMO and (\ref{eq21}) (\ref{eq23}) for MU-MIMO are all strictly concave and
monotonically increasing. At first, as $R_{\rm{s}}^{{{\rm{est}}}}$ in (\ref{eq18}) is similar with $R_{\rm{s}}({M_{\rm{a}}},{P_{\rm{t}}},W)$ in
(\ref{eq9}), the same property of strictly concave and monotonically increasing  of (\ref{eq18}) is fulfilled. About (\ref{eq21}) and
(\ref{eq23}), the proof of strictly concave and monotonically increasing is similar with the proof procedure in Theorem \ref{Theorem1}. If we
denote $g_{k,i}
> 0, i=1,\ldots,N_{{\rm{a}},k}$ as the eigenvalues of ${\bf{H}}_{{\rm{eff}},k}[n-D]{\bf{H}}_{{\rm{eff}},k}^H[n-D]$, (\ref{eq21}) and
(\ref{eq23}) can be rewritten as
\[\begin{array}{*{20}{l}}
{R_{\rm{b}}^{{\rm{est}},{\rm{low}}} = W\sum\limits_{k = 1}^{{K_{\rm{a}}}} {\sum\limits_{i = 1}^{{N_{{\rm{a}},k}}} {\log (1 +
\frac{{{P_{\rm{t}}}/{N_{\rm{s}}}}}{{{N_0}W + \sum\limits_{i = 1,i \ne k}^{{K_{\rm{a}}}} {{N_{{\rm{a}},i}}} \frac{{{P_{\rm{t}}}{\zeta
_k}}}{{{N_{\rm{s}}}}}\epsilon_{e,k}^2}}{g_{k,i}})} } }
\end{array}\]
and
\[{R_{\rm{b}}^{{\rm{est}},{\rm{upp}}} = W\sum\limits_{k = 1}^{{K_{\rm{a}}}} {\left\{ {\left[ {\sum\limits_{i = 1}^{{N_{{\rm{a}},k}}} {\log (1 +
\frac{{{P_{\rm{t}}}/{N_{\rm{s}}}}}{{{N_0}W + \sum\limits_{i = 1,i \ne k}^{{K_{\rm{a}}}} {{N_{{\rm{a}},i}}\frac{{{P_{\rm{t}}}{\zeta
_k}}}{{{N_{\rm{s}}}}}\epsilon_{e,k}^2} }}{g_{k,i}})} } \right] + ({N_{{\rm{a}},k}}/{M_{\rm{a}}}){{\log }_2}(e)} \right\}} },
\]
respectively. Calculating the first and second derivation of the above two equations, it can be proved that (\ref{eq21}) and (\ref{eq23}) are
both strictly concave and monotonically increasing in $P_{\rm{t}}$ and $W$. Therefore, based on the estimations of Proposition
\ref{Proposition0} and Proposition \ref{Proposition1}, the optimal bandwidth and transmit power can be derived at the BS.

\section{Energy Efficient Mode Switching}

\subsection{Mode Switching Based on Instant CSIT}

After getting the optimal bandwidth and transmit power for each
dedicated mode, choosing the optimal mode with optimal transmission
mode $\rm{m}^*$, optimal transmit antenna number $M_{\rm{a}}^*$,
optimal user number $K_{\rm{a}}^*$ each with optimal receive antenna
number $N_{{\rm{a}},i}^*$ is important to improve the energy
efficiency. The mode switching procedure can be described as
follows.

\underline{\emph{Energy Efficient Mode Switching Procedure}}

\emph{Step 1.} For each transmission mode $\rm{m}$ with dedicated
active transmit antenna number $M_{\rm{a}}$, active user number
$K_{\rm{a}}$ and active receive antenna number $N_{{\rm{a}},i}$,
calculate the optimal transmit power $P_{\rm{t}}^*$ and the
corresponding BPJ-EE according to the bandwidth $W^* = W_{\rm{max}}$
and capacity estimation based on Proposition \ref{Proposition0} and
Proposition \ref{Proposition1}.

\emph{Step 2.} Choose the optimal transmission mode $\rm{m}^*$ with
optimal $M_{\rm{a}}^*$, $K_{\rm{a}}^*$ and $N_{{\rm{a}},i}^*$ with
the maximum BPJ-EE. $\Box$

The above procedure is based on the  instant CSIT. As we know, there are two main schemes to choose the optimal mode  in the spectral efficient
multimode transmission systems. The one is based on the instant CSIT \cite{Shen,Rchen1,Rchen2}, while the other is based on the ergodic capacity
\cite{Zhang1,Zhang2,Zhang3}. The ergodic capacity based mode switching can be performed off line and can provide more guidance on the preferred
mode under given scenarios. If applying the ergodic capacity of each mode in the energy efficient mode switching, similar benefits can be
exploited. The next subsection will present the approximation of ergodic capacity and propose the ergodic capacity based mode switching.

\subsection{Mode Switching Based on the Ergodic Capacity}

Firstly, the ergodic capacity of each mode need to be developed. The
following lemma gives the asymptotic result of the point to point
MIMO channel with full CSIT when $M_{\rm{a}} \geq N_{\rm{a}}$.

\begin{Lemma}\label{Lemma3}
\cite{Zhang2,Rapajic} For a point to point channel when $M_{\rm{a}}
\geq N_{\rm{a}}$, denote $\beta = \frac{M_{\rm{a}}}{N_{\rm{a}}}$ and
$\gamma = \frac{P_{\rm{t}}\zeta _k }{N_0W}$. The capacity is
approximated as
\begin{equation} \label{eq25}
\begin{array}{l}
R_{\rm{s}}^{\rm{appro}} \approx W{\mathcal
{C}}_{{\rm{iso}}}(\beta,\beta\gamma)
\end{array}
\end{equation}
in which ${\mathcal{C}}_{{\rm{iso}}}$ is the asymptotic spectral efficiency of the point to point channel, and ${\mathcal{C}}_{{\rm{iso}}}$ can
be denoted as
\begin{equation} \label{eq26}
\begin{array}{l}
\frac{{{\mathcal {C}}}_{\rm{iso}}(\beta,\gamma)}{N_{\rm{a}}} =
\log_2\left[1+\gamma-\mathcal {F}(\beta,\frac{\gamma}{\beta})\right]
\\ + \beta\log_2\left[1+\frac{\gamma}{\beta}-\mathcal
{F}(\beta,\frac{\gamma}{\beta})\right] -
\beta\frac{\log_2(e)}{\gamma} \mathcal
{F}(\beta,\frac{\gamma}{\beta})
\end{array}
\end{equation}
with
\[
{\mathcal {F}}(x,y) =
\frac{1}{4}\left[\sqrt{1+y(1+\sqrt{x})^2}-\sqrt{1+y(1-\sqrt{x})^2}\right]^2.
\]
\end{Lemma}

As SVD is applied in the SU-MIMO systems, the transmission is
aligned with the maximum $N_{\rm{s}}$ singular vectors. When
$M_{\rm{a}} < N_{\rm{a}}$, the achievable capacity approximation is
modified as
\begin{equation} \label{eq27}
\begin{array}{l}
R_{\rm{s}}^{\rm{appro}} \approx W\mathcal
{C}_{\rm{iso}}({\hat{\beta}},\hat{\beta}{\gamma}),
\end{array}
\end{equation}
where $\hat{\beta} = \frac{1}{\beta} =
\frac{N_{\rm{a}}}{M_{\rm{a}}}$.

Therefore, according Proposition \ref{Proposition0}, the following
proposition can be get directly.

\begin{Proposition}\label{Proposition2}
The ergodic capacity of SU-MIMO with SVD is estimated by:
\begin{equation} \label{eq18_1}
\begin{array}{l}
R_{\rm{s}}^{{{\rm{Ergodic}}}} =R_{\rm{s}}^{\rm{appro}}.
\end{array}
\end{equation}
\end{Proposition}

Although Zhang et al. give another accurate approximation in
\cite{Zhang2} for the MU-MIMO systems with BD, it is only applicable
in the scenario in which
$\sum\limits_{i=1}^{K_{\rm{a}}}N_{{\rm{a}},i} = {M_{\rm{a}}}$. We
develope the ergodic capacity estimation with BD based on
Proposition \ref{Proposition1}.

As ${\bf{T}}_k[n-D]$ is designed to null the interuser interference,
it is a unitary matrix independent of ${\bf{H}}_k[n-D]$. So
${\bf{H}}_k[n-D]{\bf{T}}_k[n-D]$ is also a zero-mean complex
Gaussian matrix with dimension $N_{{\rm{a}},k} \times M_{{\rm{a}},k}
$, where $M_{{\rm{a}},k} = {M_{\rm{a}}} - \sum\limits_{i=1,i\neq
k}^{{K_{\rm{a}}}}{N_{{\rm{a}},i}}$. The effective channel matrix of
user $k$ can be treated as a SU-MIMO channel with transmit antenna
number $M_{{\rm{a}},k}$ and receive antenna number $N_{{\rm{a}},k}$.
Combining Proposition \ref{Proposition0}, Proposition
\ref{Proposition1} and Proposition \ref{Proposition2}, we have the
following Proposition.

\begin{Proposition}\label{Proposition3}
The lower bound of the ergodic capacity estimation of MU-MIMO with
BD is given by
\begin{equation} \label{eq28}
\begin{array}{l}
R_{\rm{b}}^{{\rm{Ergodic-low}}}  \approx W\sum \limits
_{k=1}^{K_{\rm{a}}} {\mathcal
{C}}_{\rm{iso}}({\hat{\beta}}_k,{\hat{\beta}}_k{\hat{\gamma}}_k),
\end{array}
\end{equation}
while the upper bound of the ergodic capacity estimation of MU-MIMO
with BD is given by
\begin{equation} \label{eq29}
\begin{array}{l}
R_{\rm{b}}^{{\rm{Ergodic-upp}}} \approx W\sum \limits
_{k=1}^{K_{\rm{a}}} \left[{\mathcal
{C}}_{\rm{iso}}(\hat{\beta}_k,\hat{\beta}_k\hat{\gamma}_k) +
\frac{1}{\hat{\beta}_k}\log_2(e)\right],
\end{array}
\end{equation}
where \[\hat{\beta}_k = M_{{\rm{a}},k} / N_{{\rm{a}},k},\]
\[\hat{\gamma}_k = \frac{P_{\rm{t}}\zeta_k }{{N_0W+\sum\limits_{i=1,i\neq
k}^{{K_{\rm{a}}}}{N_{{\rm{a}},i}}\frac{P_{\rm{t}}\zeta_k}{N_{\rm{s}}}\epsilon_{e,k}^2}}.\]
\end{Proposition}

For comparison, the ergodic capacity lower bound based on (\ref{eq20}) is also considered. As shown in (\ref{eq19}), the expectation can be
denoted as
\[\mathbb{E}(R_{\rm{b}}^P - R_{\rm{b}}^D) \leq \mathbb{E}( \triangle R_{\rm{b}}^{\rm{upp}}). \]
As $\triangle R_{\rm{b}}^{\rm{upp}}$ is a constant, we have $\mathbb{E}( \triangle R_{\rm{b}}^{\rm{upp}}) = \triangle R_{\rm{b}}^{\rm{upp}}$,
and then
\begin{equation} \label{eq30_before}
\begin{array}{l}
\mathbb{E}(R_{\rm{b}}^P) - \mathbb{E}(R_{\rm{b}}^D) \leq \triangle R_{\rm{b}}^{\rm{upp}}.
\end{array}
\end{equation}
Therefore, the lower bound estimation in (\ref{eq20}) can also be applied to the ergodic capacity case. As the expectation of (\ref{eq19_10})
can be denoted as \cite{Zhang2}
\begin{equation} \label{eq30_before1}
\begin{array}{l}
\mathbb{E}(R_{\rm{b}}^{{\rm{est}},P}) = W\sum \limits _{k=1}^{K_{\rm{a}}} {\mathcal {C}}_{\rm{iso}}({\hat{\beta}}_k,{\hat{\beta}}_k{{\gamma}}),
\end{array}
\end{equation}
the low bound ergodic capacity estimation can be denoted as
\begin{equation} \label{eq30}
\begin{array}{l}
R_{\rm{b}}^{{\rm{Ergodic-Zhang}}} \approx W\sum \limits _{k=1}^{K_{\rm{a}}} {\mathcal {C}}_{\rm{iso}}({\hat{\beta}}_k,{\hat{\beta}}_k{{\gamma}})
- \triangle R_{\rm{b}}^{\rm{upp}}.
\end{array}
\end{equation}

After getting the ergodic capacity of each mode, the ergodic
capacity based mode switching algorithm can be summarized as
follows.

\underline{\emph{Ergodic Capacity Based Energy Efficient Mode
Switching}}

\emph{Step 1.} For each transmission mode $\rm{m}$ with dedicated
 $M_{\rm{a}}$,
$K_{\rm{a}}$ and  $N_{{\rm{a}},i}$, calculate the optimal transmit
power $P_{\rm{t}}^*$ and the corresponding BPJ-EE according to the
bandwidth $W^* = W_{\rm{max}}$ and ergodic capacity estimation based
on Proposition \ref{Proposition2} and Proposition
\ref{Proposition3}.

\emph{Step 2.} Choose the optimal $\rm{m}^*$ with optimal
$M_{\rm{a}}^*$, $K_{\rm{a}}^*$ and $N_{{\rm{a}},i}^*$ with the
maximum BPJ-EE. $\Box$

According to the ergodic capacity based mode switching scheme, the operation mode under dedicated scenarios can be determined in advance. Saving
a lookup table at the BS according to the ergodic capacity based mode switching, the optimal mode can be chosen simply according to the
application scenarios. The performance and the preferred mode in a given scenario will be shown in the next section.

\section{Simulation Results}

\begin{figure*}[!t]
\begin{center}
\includegraphics[height = 1.55in] {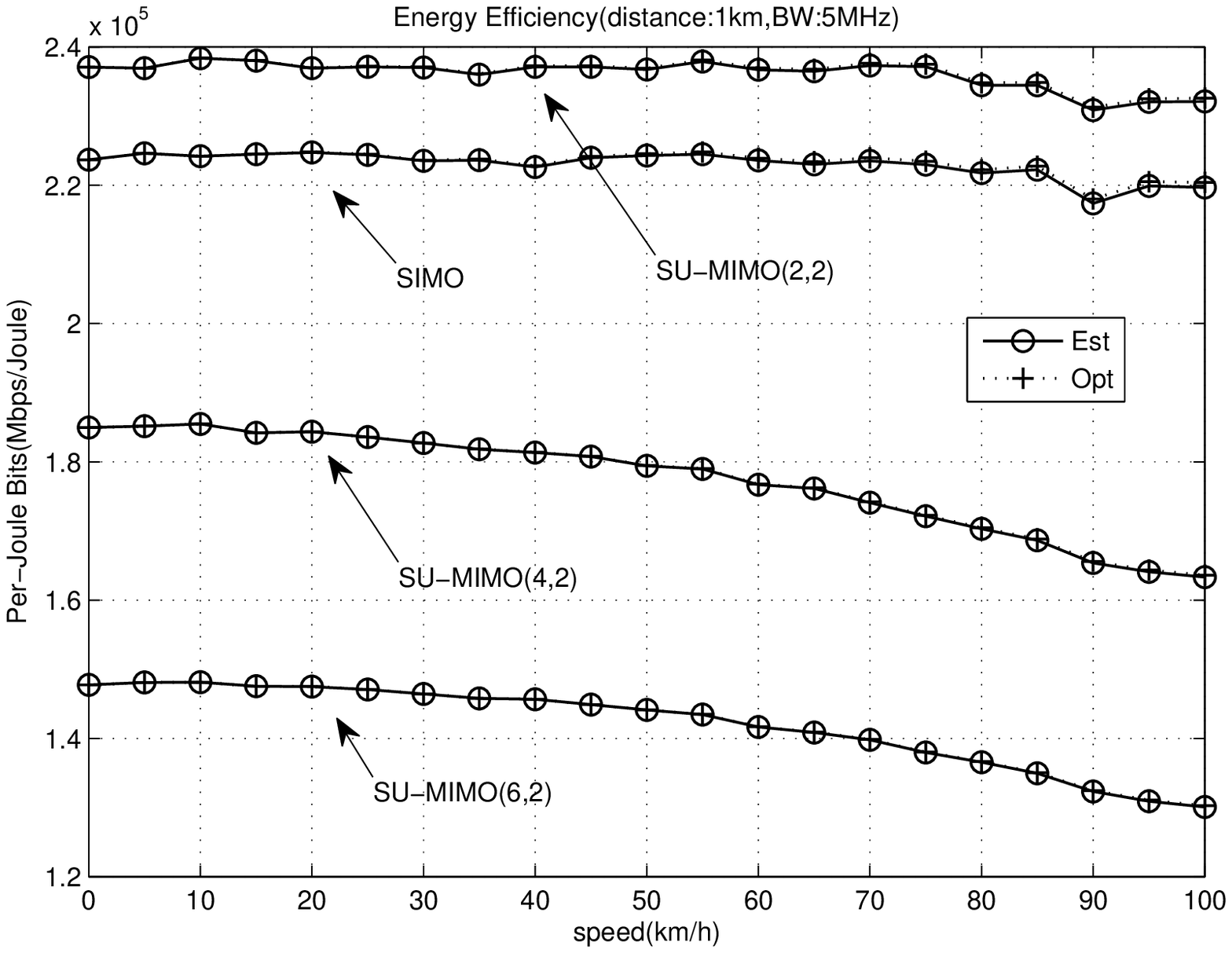}
\includegraphics[height = 1.55in] {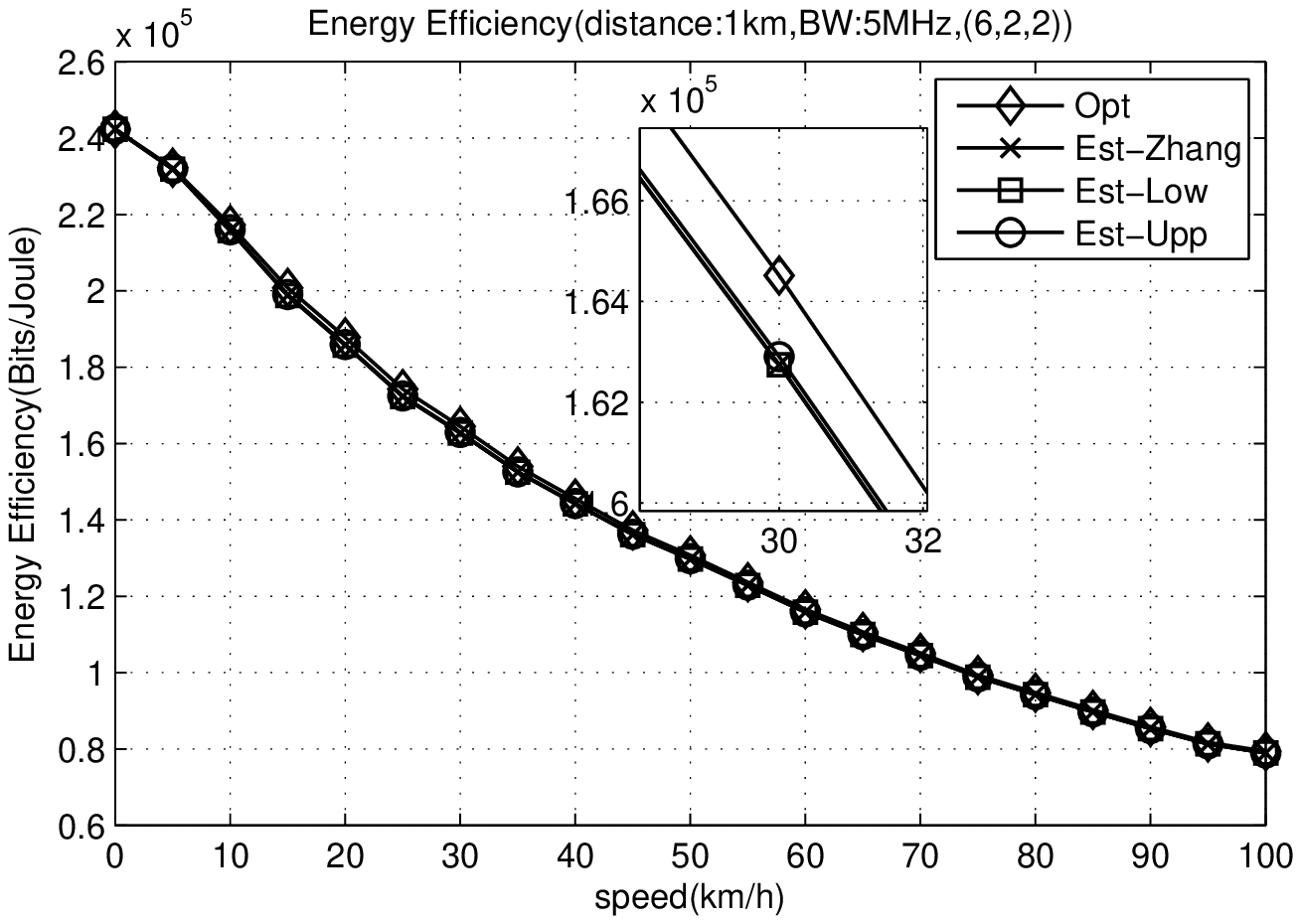}
\includegraphics[height = 1.55in] {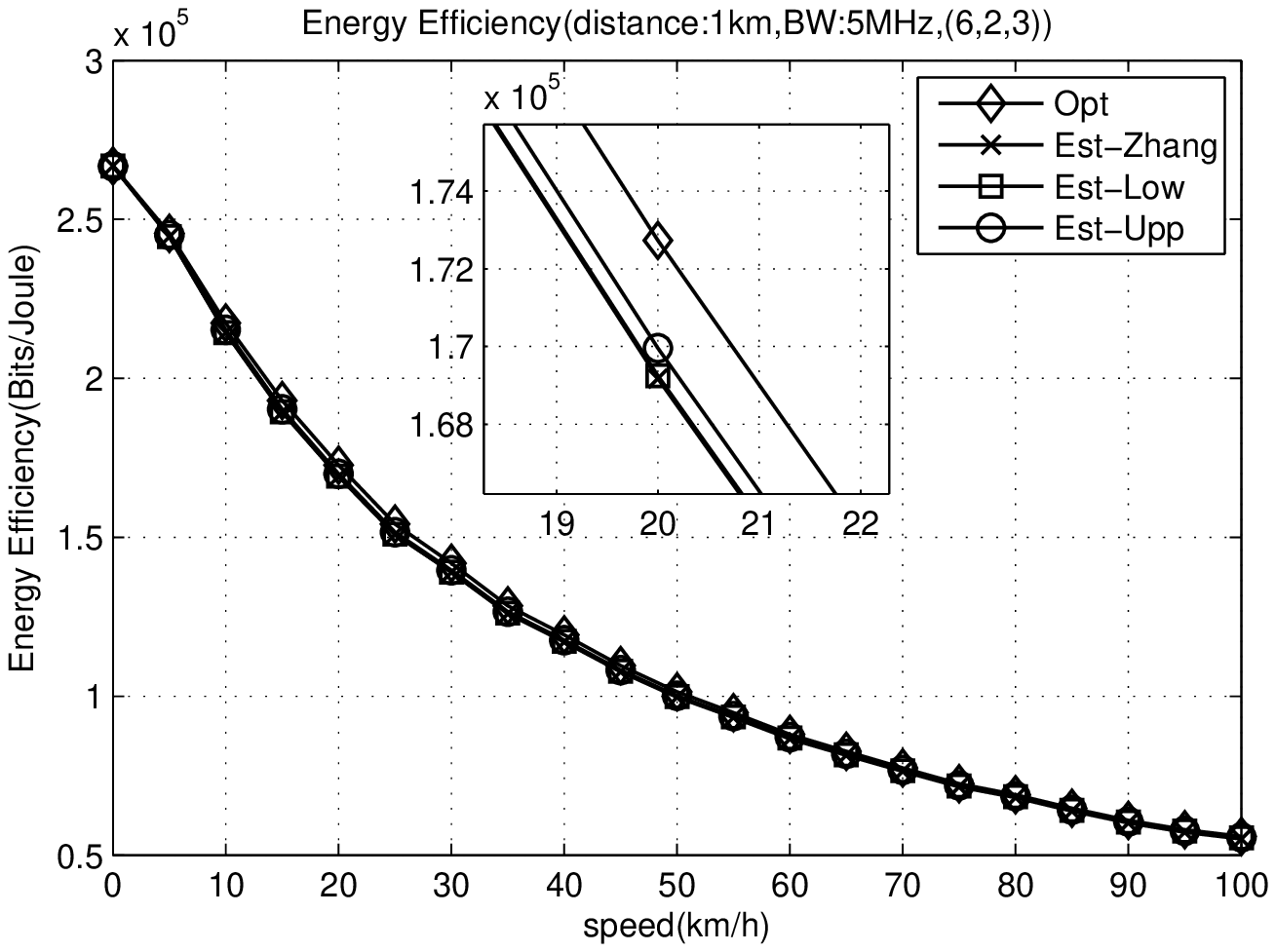}
\end{center}
\caption{The effect of capacity estimation on the energy efficiency
of SU-MIMO and MU-MIMO under different speed.} \label{fig3}
\end{figure*}

\begin{figure*}[!t]
\begin{center}
\includegraphics[height = 1.55in] {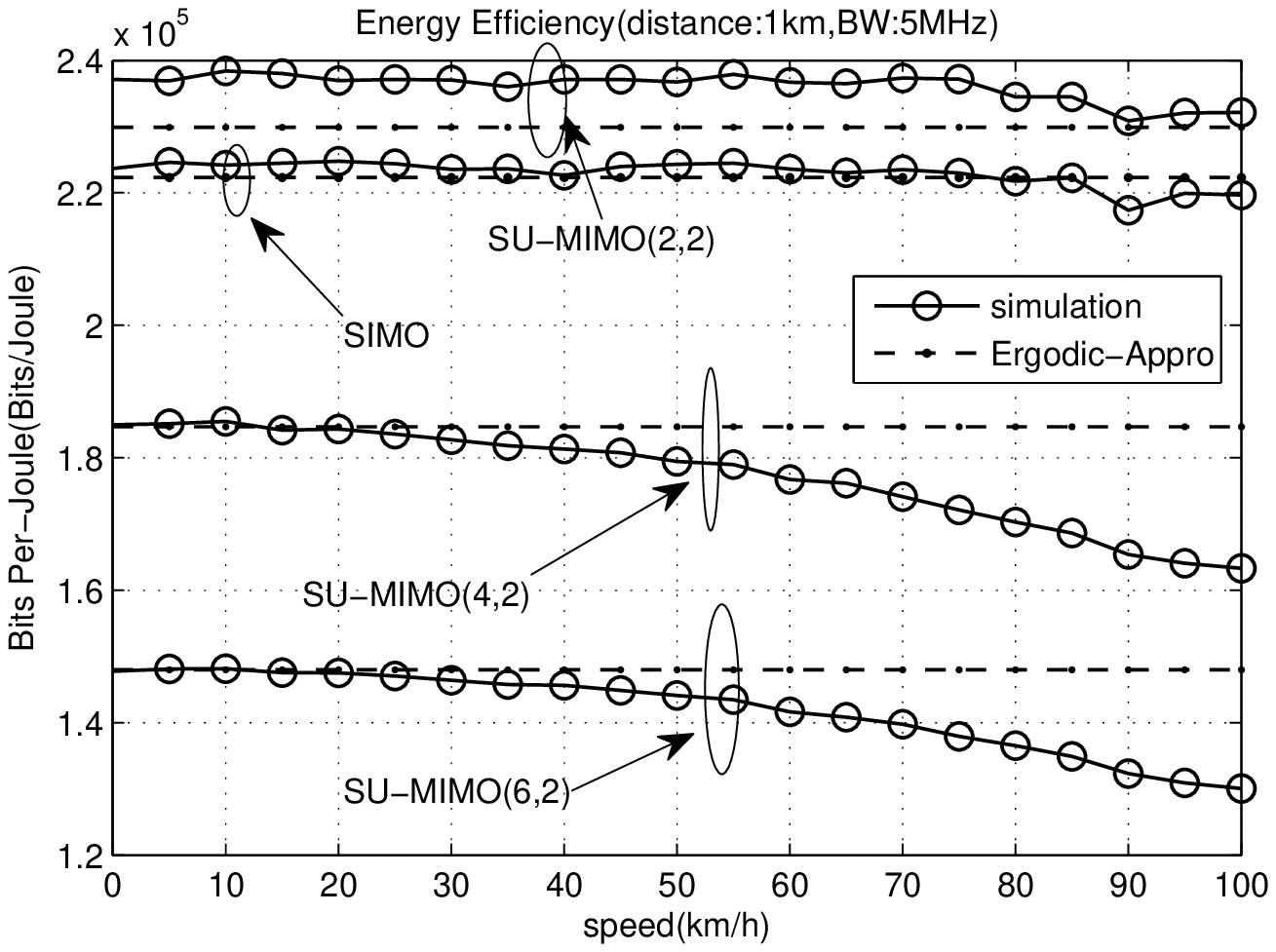}
\includegraphics[height = 1.55in] {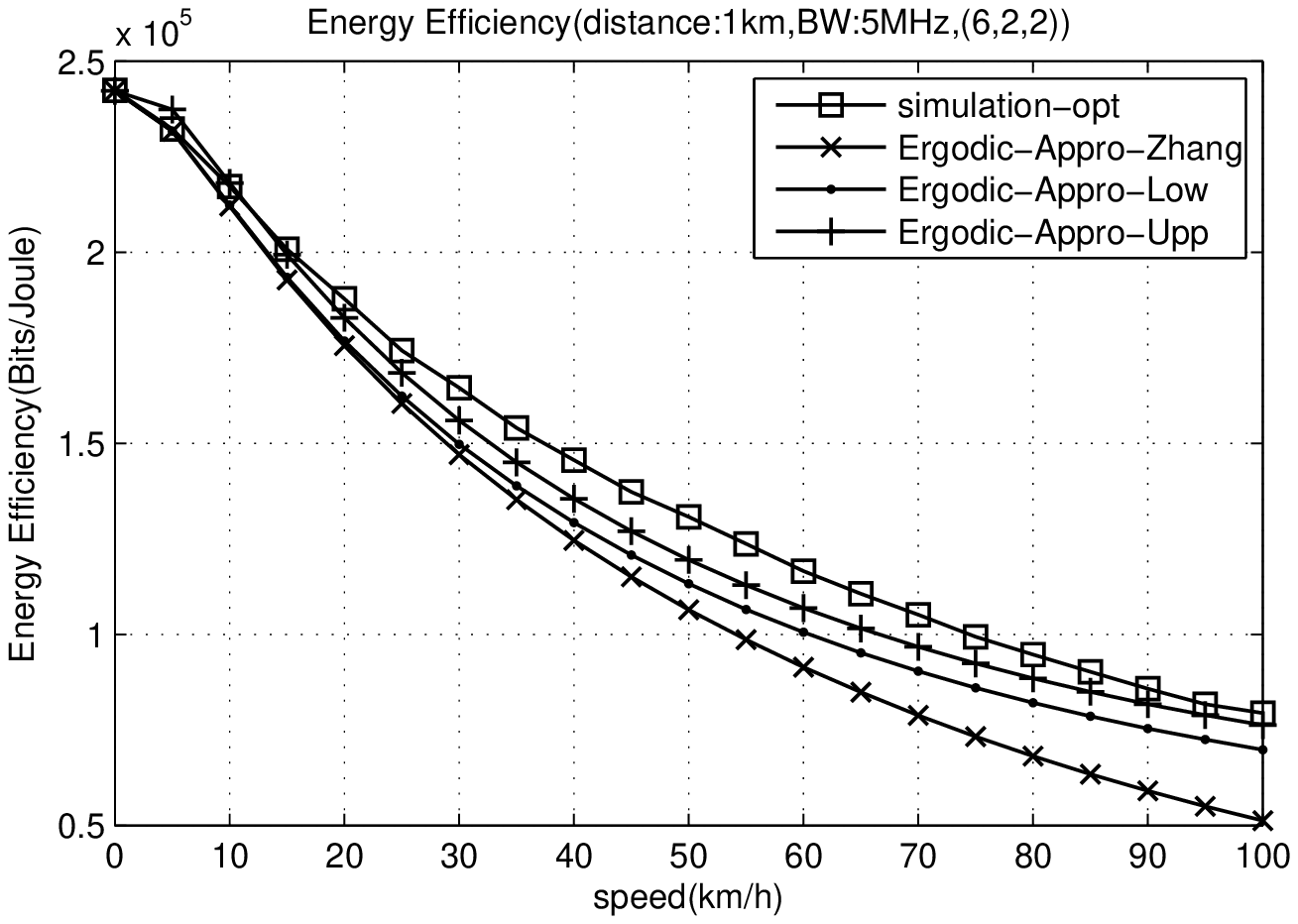}
\includegraphics[height = 1.55in] {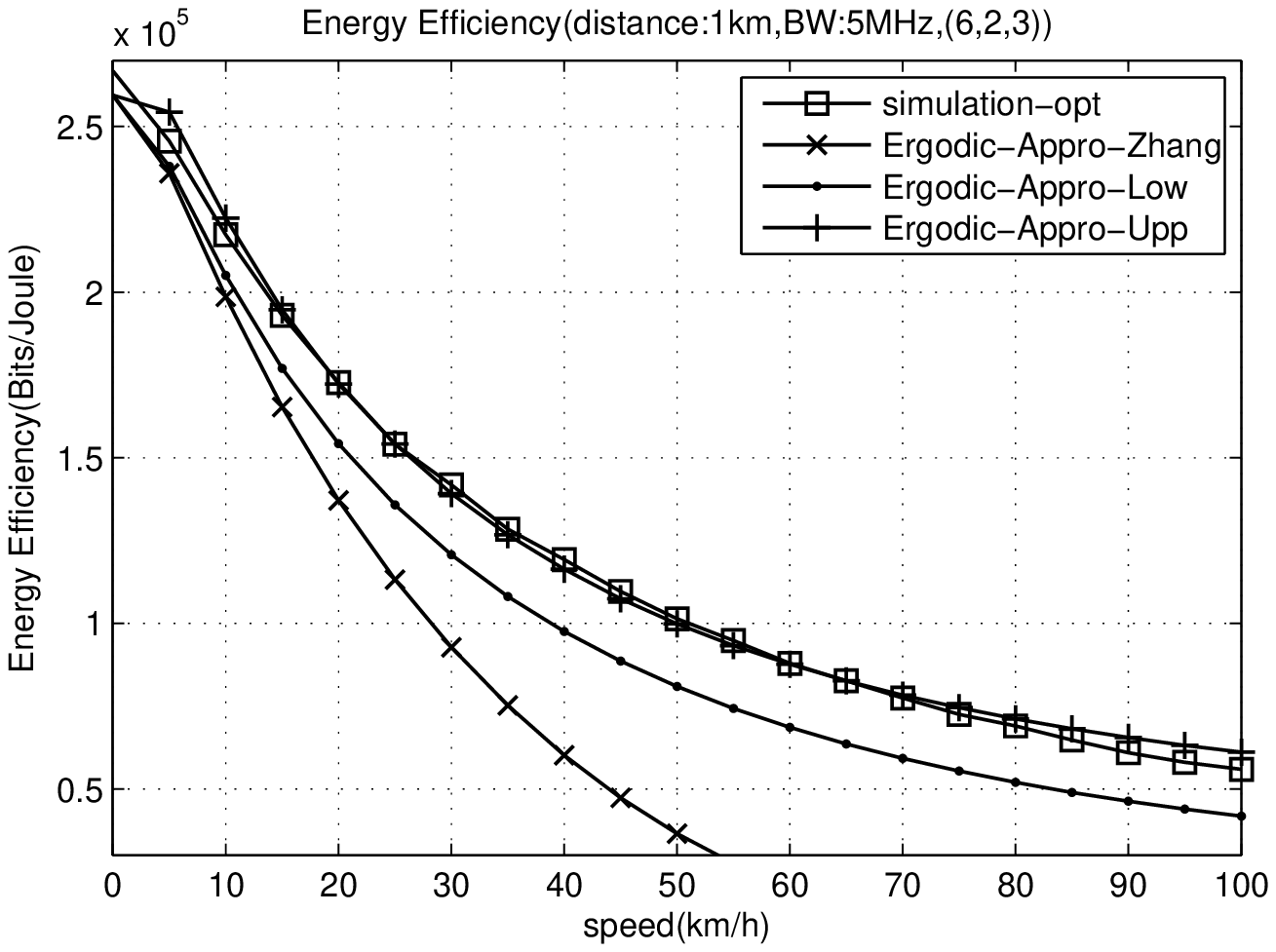}
\end{center}
\caption{Comparison of energy efficiency based on ergodic capacity
and instant capacity with SU-MIMO and MU-MIMO} \label{fig6}
\end{figure*}

\begin{figure*}[!t]
\begin{center}
\includegraphics[height = 1.55in] {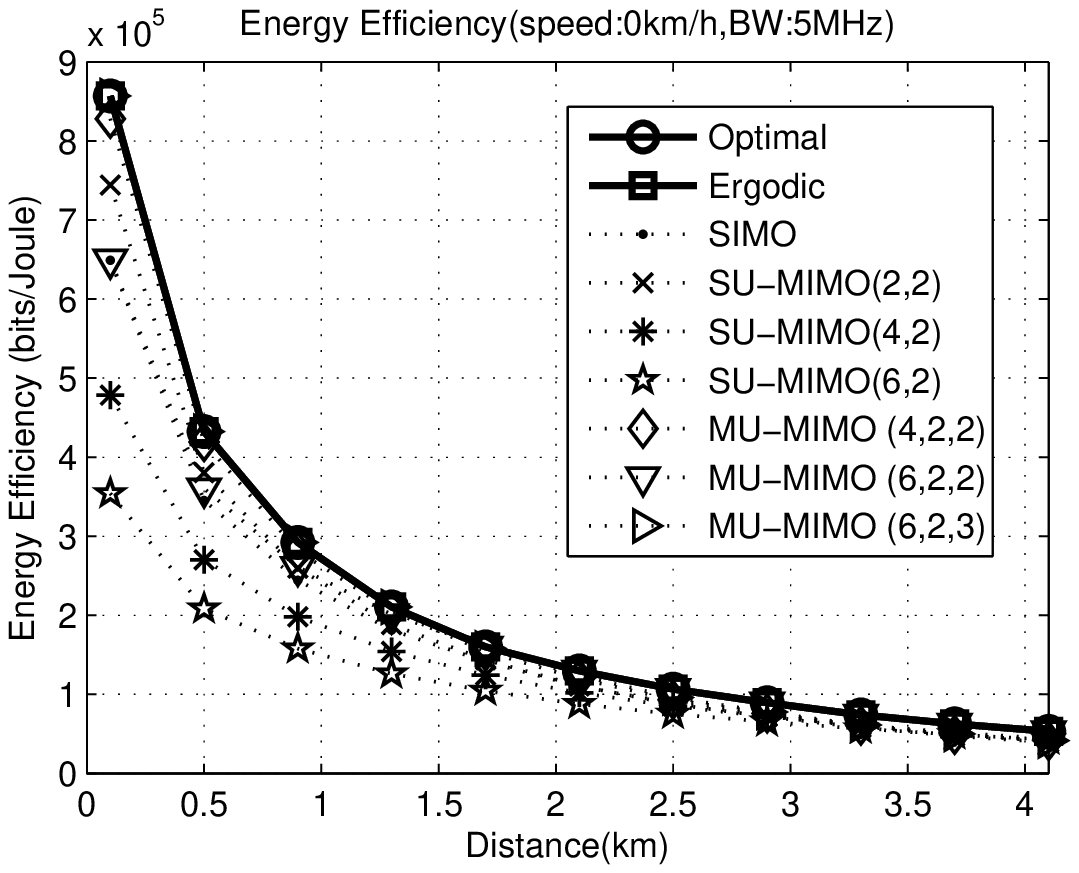}
\includegraphics[height = 1.55in] {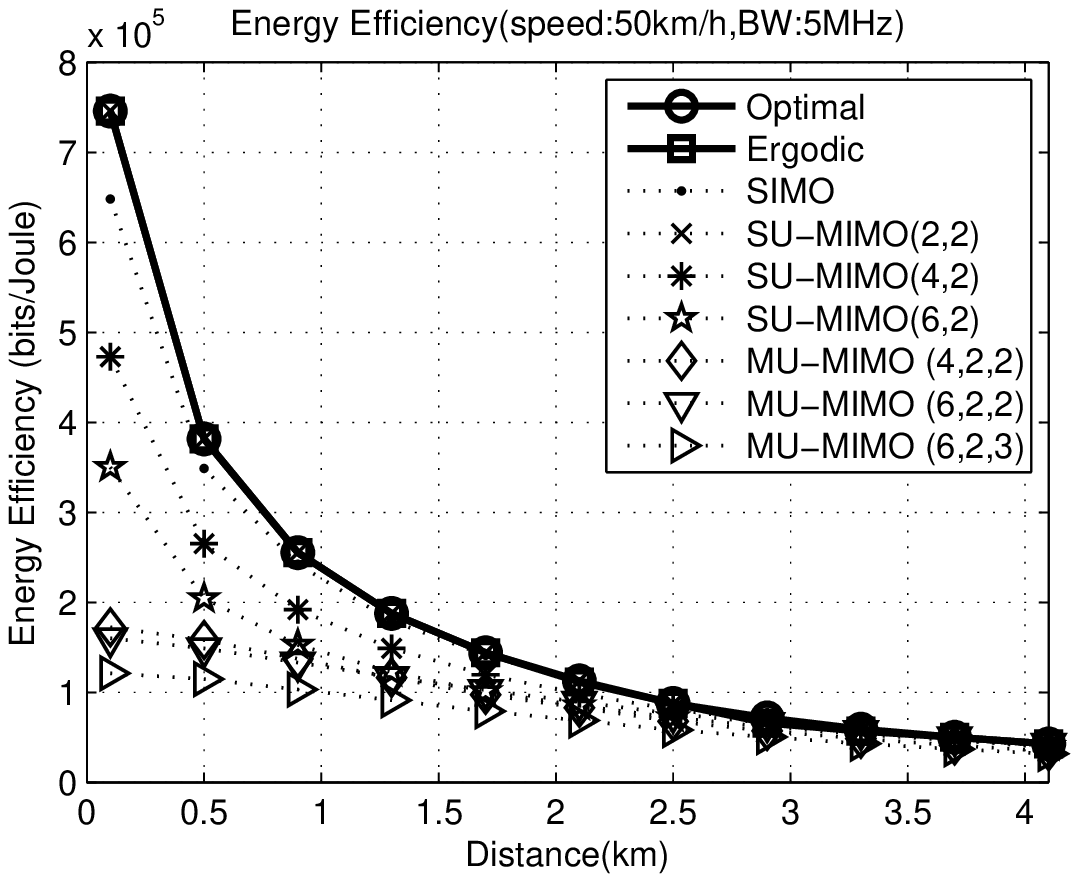}
\includegraphics[height = 1.55in] {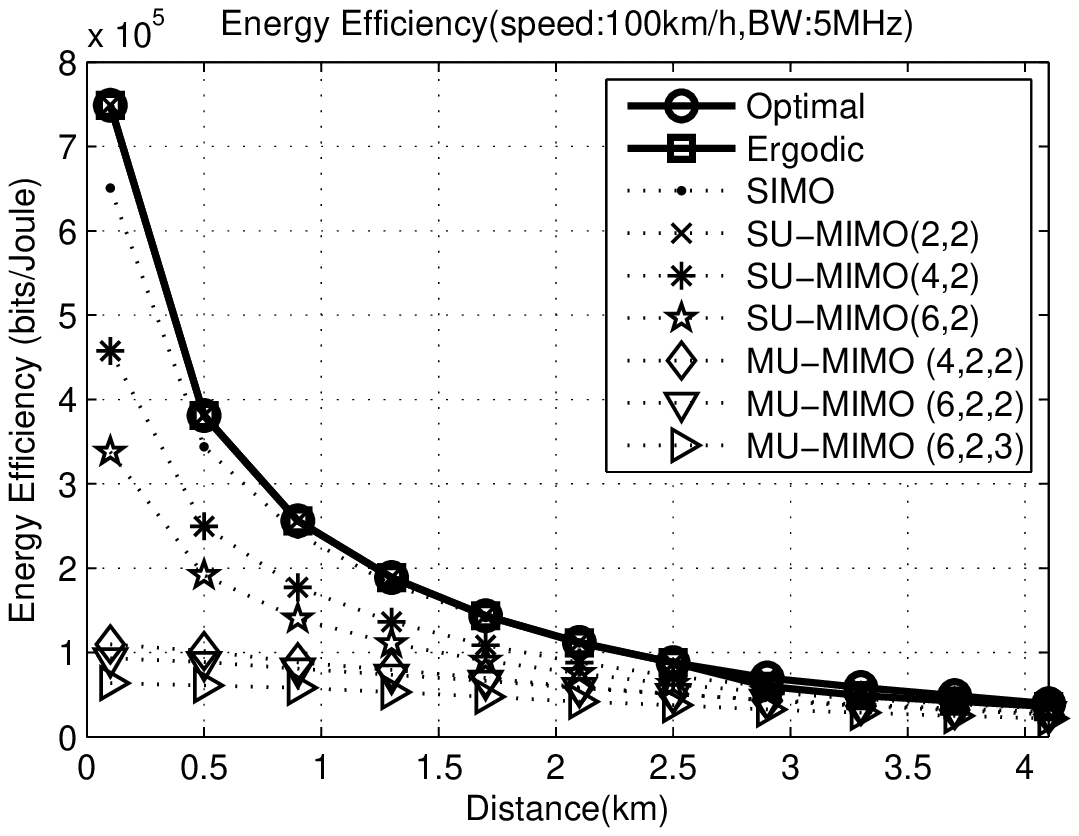}
\end{center}
\caption{Performance of mode switching} \label{fig7}
\end{figure*}

This section provides the simulation results. In the simulation, $M=6$, $N=2$ and $K=3$. All users are assumed to be homogeneous with the same
distance and moving speed. Only pathloss is considered for the large scale fading model and the pathloss model is set as $128.1+37.6
\log_{10}d_k$ dB ($d_k$ in kilometers). Carrier frequency is set as $2{\rm{GHz}}$ and $D=1{\rm{ms}}$. Noise density is $N_0 =
-174{\rm{dBm/Hz}}$. The power model is modified according to \cite{Arnold}, which is set as $\eta = 0.38$, $P_{\rm{cir}} = 66.4 {\rm{W}}$,
$P_{\rm{Sta}} = 36.4 {\rm{W}}$, $p_{\rm{sp,bw}} = 3.32 \mu{\rm{W/Hz}}$ and $p_{\rm{ac,bw}} = 1.82 \mu{\rm{W/Hz}}$. $W_{\max} = 5{\rm{MHz}}$. For
simplification, "SU-MIMO ($M_{\rm{a}}$,$N_{\rm{a}}$)" denotes SU-MIMO mode with $M_{\rm{a}}$ active transmit antennas and $N_{\rm{a}}$ active
receive antennas, "SIMO" denotes SU-MIMO mode with one active transmit antennas and $N$ active receive antennas and "MU-MIMO
($M_{\rm{a}}$,$N_{\rm{a}}$,$K_{\rm{a}}$)" denotes MU-MIMO mode with $M_{\rm{a}}$ active transmit antennas and $K_{\rm{a}}$ users each
$N_{\rm{a}}$ active receive antennas. Seven transmission modes are considered in the simulation, i.e. SIMO, SU-MIMO (2,2), SU-MIMO (4,2),
SU-MIMO (6,2), MU-MIMO (4,2,2), MU-MIMO (6,2,2), MU-MIMO (6,2,3). In the simulation, the solution of (\ref{eq5_1})(\ref{eq5_3}) and (\ref{eq16})
is derived by the Newton's method, as the close-form solution is difficult to get.

Fig. \ref{fig3} depicts the effect of capacity estimation on the optimal BPJ-EE under different moving speed. The optimal estimation means that
the BS knows the channel error during calculating $P_{\rm{t}}^*$ and the precoding is still based on the delayed CSIT. In the left figure,
SU-MIMO is plotted. The performance of capacity estimation and the optimal estimation are almost the same, which indicates that the capacity
estimation of the SU-MIMO systems is robust to the delayed CSIT. Another observation is that the BPJ-EE is nearly constant as the moving speed
is increasing for SIMO and SU-MIMO(2,2), while it is decreasing for SU-MIMO(4,2) and SU-MIMO(6,2). The reason can be illustrated as follows. The
precoding at the BS can not completely align with the singular vectors of the channel matrix under the imperfect CSIT. But when the transmit
antenna number is equal to or greater than the receive antenna number, the receiver can perform detection to get the whole channel matrix's
degree of freedom. However, when the transmit antennas are less than the receive antenna, the receiver cannot get the whole degree of freedom
only through detection, so the degree of freedom loss occurs. The center and right figures show us the effect of capacity estimation with
MU-MIMO modes. The three estimation schemes all track the effect of imperfect CSIT. From the amplified sub-figures, the upper bound capacity
estimation is the closest one to the optimal estimation. It indicates that the upper bound capacity estimation is the best one in the BD scheme.
Moreover, we can see that BPJ-EE of the BD scheme decreases seriously due to the imperfect CSIT caused inter-user interference.

Fig. \ref{fig6} compares the BPJ-EE derived by ergodic capacity estimation schemes and the one by simulations. The left figure demonstrates the
SU-MIMO modes. The estimation of SIMO, SU-MIMO(4,2) and SU-MIMO(6,2) is accurate when the moving speed is low. But when the speed is increasing,
the ergodic capacity estimation of SU-MIMO(4,2) and SU-MIMO(6,2) can not track the decrease of BPJ-EE. There also exists a gap between the
ergodic capacity estimation and the simulation in the SIMO mode. Although the mismatching exists, the ergodic capacity based mode switching can
always match the optimal mode, which will be shown in the next figure. For the MU-MIMO modes, the two lower bound ergodic capacity estimation
schemes mismatch the simulation more than the upper bound estimation scheme. That is because the lower bound estimations cause BPJ-EE decreasing
twice. Firstly, the derived transmit power would mismatch the exactly accurate transmit power because the derivation is based on a bound and
this transmit power mismatch will make the BPJ-EE decrease compared with the simulation. Secondly, the lower bound estimation use a lower bound
formula to calculate the estimated BPJ-EE under the derived transmit power which will make the BPJ-EE decrease again. Nevertheless, the upper
bound estimation has the opposite impact on the BPJ-EE estimation during the above two steps, so it matches the simulation much better.
According to Fig. \ref{fig3} and Fig. \ref{fig6}, the upper bound estimation is the best estimation scheme for the MU-MIMO mode. Therefore,
during the ergodic capacity based mode switching, the upper bound estimation is applied.

Fig. \ref{fig7} depicts the BPJ-EE performance of mode switching. For comparison, the optimal mode with instant CSIT (`Optimal') is also
plotted. The mode switching can improve the energy efficiency significantly and the ergodic capacity based mode switching can always track the
optimal mode. The performance of ergodic capacity based switching is nearly the same as the optimal one. Through the simulation, the ergodic
capacity based mode switching is a promising way to choose the most energy efficient transmission mode.

\begin{figure}[!h]
\begin{center}
\includegraphics[height = 2.5in] {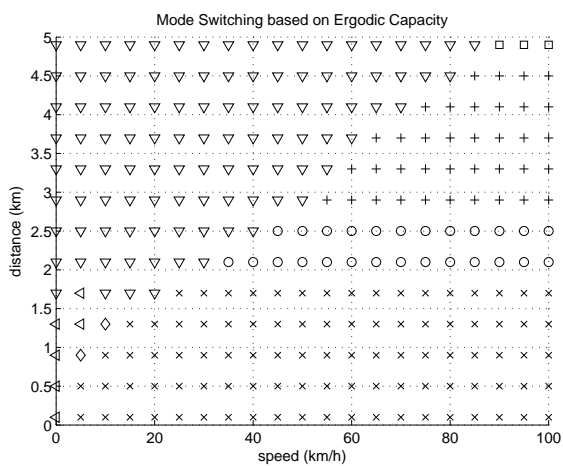}
\end{center}
\caption{Optimal mode under different scenario. $\circ$: SIMO,
$\times$: SU-MIMO (2,2), $+$: SU-MIMO (4,2),$\Box$: SU-MIMO
(6,2),$\lozenge$: MU-MIMO (4,2,2),$\nabla$: MU-MIMO
(6,2,2),$\vartriangleleft$: MU-MIMO (6,2,3). } \label{fig1}
\end{figure}

Fig. \ref{fig1} demonstrates the preferred transmission mode under given scenarios. The optimal mode under different moving speed and distance
is depicted. This figure provides insights on the PC power/dynamic power/static power tradeoff and the multiplexing gain/inter-user interference
compromise. When the moving speed is low, MU-MIMO modes are preferred and vice versa. This result is similar to the spectral efficient mode
switching in \cite{Zhang1,Zhang2,Zhang3,Xu2}. Inter-user interference is small when the moving speed is low, so higher multiplexing gain of
MU-MIMO benefits. When the moving speed is high, the inter-user interference with MU-MIMO becomes significant, so SU-MIMO which can totally
avoid the interference is preferred. Let us focus on the effect of distance on the mode under high moving speed case then. When distance is less
than 1.7km, SU-MIMO (2,2) is the optimal one, while the distance is equal to 2.1 km and 2.5km, the SIMO mode is suggested. When the distance is
larger than 2.5km, the active transmit antenna number increases as the distance increases. The reason of the preferred mode variation can be
explained as follows. As the total power can be divided into PC power, transmit antenna number related power "Dyn-I" and "Dyn-III" and transmit
antenna number independent power "Dyn-II" and static power. The first and third part divided by capacity would increase as the active number
increases, while the second part is opposite. In the long distance scenario, the first part will dominates the total power and then more active
antenna number is preferred. In the short and medium distance scenario, the second and third part dominate the total power and the tradeoff
between the two parts should be met. Above all, the above mode switching trends of Fig. \ref{fig1} externalize the two tradeoffs.

\section{Conclusion}
This paper discusses the energy efficiency maximizing problem in the downlink MIMO systems. The optimal bandwidth and transmit power are derived
for each dedicated mode with constant system parameters, i.e. fixed transmission scheme, fixed active transmit/receive antenna number and fixed
active user number. During the derivation, the capacity estimation mechanism is presented and several accurate capacity estimation strategies
are proposed to predict the capacity with imperfect CSIT. Based on the optimal derivation, ergodic capacity based mode switching is proposed to
choose the most energy efficient system parameters. This method is promising according to the simulation results and provides guidance on the
preferred mode over given scenarios.

\appendices
\section{Proof of Lemma \ref{lemma0}}\label{A_1}
{\bf{Proof:}} The proof of the above lemma is motivated by
\cite{Miao1}. Denote the inverse function of $y=f(x)$ as $x=g(y)$,
then $x^* = \arg \max_x \frac{f(x)}{ax+b} = \arg \max_{g(y)}
\frac{y}{ag(y)+b}$. Denote $y^*=f(x^*)$. Since $f(x)$ is
monotonically increasing, $y^* = \arg \max_{y} \frac{y}{ag(y)+b}$.
According to \cite{Miao1}, there exists a unique globally optimal
$y^*$ given by
\begin{equation} \label{eq5_2}
\begin{array}{l}
y^{*}  = \frac{b+ag(y^{*})}{ag^{'}(y^{*})}
\end{array}
\end{equation}
if $g(y)$ is strictly convex and monotonically increasing. (\ref{eq5_2}) is fulfilled since the inverse function of $g(y)$, i.e. $f(x)$ is
strictly concave and monotonically increasing. Taking $g^{'}(y) = \frac{1}{f^{'}(x)}$ and $f(x) = y$ into (\ref{eq5_2}), we can get
(\ref{eq5_1}).

\section{Proof of Theorem \ref{Theorem0}}\label{A_2}

{\bf{Proof:}} The first part can be proved according to Lemma \ref{lemma0}. Calculating the first and second derivation of $R(W)$ based on
(\ref{eq9}), (\ref{eq12}) and (\ref{eq13}), we can see that $R(W)$ of both SVD and BD mode is strictly concave and monotonically increasing as a
function of $W$. The optimal $W^{*}$ can be get through (\ref{eq5_1}), which is given by (\ref{eq5_3}).

Look at the second part. Taking $P_{\rm{t}} = p_{\rm{t}} W$ into
(\ref{eq9}), (\ref{eq12}) and (\ref{eq13}), the capacity is
$R(P_{\rm{t}},W) = W {\hat{R}}_{\rm{m}}(p_{\rm{t}})$, where
${\hat{R}}(p_{\rm{t}})$ is independent of $W$. We have that
\begin{equation} \label{eq5_4}
\begin{array}{l}
{\xi}  = \frac{W {\hat{R}}(p_{\rm{t}})}{{ \left(M_{\rm{a}} p_{\rm{sp,bw}} + p_{\rm{ac,bw}} \right) W + M_{\rm{a}} P_{\rm{cir}} + P_{\rm{PC}}+
P_{\rm{Sta}}}}.
\end{array}
\end{equation}
The second part is verified.

\section{Proof of Theorem \ref{Theorem1}}\label{A_3}

{\bf{Proof:}} According to Lemma \ref{lemma0}, the above theorem can be verified if we prove that $R_{\rm{m}}(P_{\rm{t}})$ is strictly concave
and monotonically increasing for both SVD and BD. It is obvious that the capacity of SVD and BD with perfect CSIT is strictly concave and
monotonically increasing based on (\ref{eq9}) and (\ref{eq12}). If the capacity of BD with imperfect CSIT can also be proved to be strictly
concave and monotonically increasing, Theorem \ref{Theorem1} can be proved.

Denoting ${\bf{A}}_k = {\bf{E}}_k[n]\left[ \sum \limits_{i\neq k} {\bf{T}}^{(D)}_i[n] {\bf{T}}_i^{(D)H} [n] \right] {\bf{E}}_k^H [n]$, then
rewrite (\ref{eq13}) as follows:
\begin{equation} \label{eq16_1}
\begin{array}{l}
R_{\rm{b}}^D ({P_{\rm{t}}}) =W\sum\limits_{k=1}^{K_{\rm{a}}}\left \{\log\det\left( {\bf{R}}_k[n] +
\frac{P_{t}}{{N_{\rm{s}}}}\hat{\bf{H}}_{{\rm{eff}},k}[n]\hat{\bf{H}}_{{\rm{eff}},k}^H
[n] \right)  - \log\det{\bf{R}}_k[n] \right\}\\
=W\sum\limits_{k=1}^{K_{\rm{a}}}\left \{\log\det\left[ {\bf{I}} + \frac{P_{\rm{t}}}{N_0W N{\rm{s}}} \left({\bf{A}}_k +
\hat{\bf{H}}_{{\rm{eff}},k}[n]\hat{\bf{H}}_{{\rm{eff}},k}^H [n]\right) \right] - \log\det\left({\bf{I}} + \frac{P_{\rm{t}}}{N_0W N{\rm{s}}}
{\bf{A}}_k \right) \right\} \\
=W\sum\limits_{k=1}^{K_{\rm{a}}} \sum\limits_{i=1}^{N_{{\rm{a}},k}} \left \{\log(1+\frac{P_{\rm{t}}}{N_0W N{\rm{s}}} c_{k,i}) -
\log(1+\frac{P_{\rm{t}}}{N_0W N{\rm{s}}} g_{k,i})\right\}.
\end{array}
\end{equation}
$c_{k,i}$ and $g_{k,i}$ are the eigenvalue of ${\bf{A}}_k + \hat{\bf{H}}_{{\rm{eff}},k}[n]\hat{\bf{H}}_{{\rm{eff}},k}^H [n]$ and ${\bf{A}}_k$,
respectively. Sorting $c_{k,i}$ and $g_{k,i}$ as $c_{k,1} \ge \ldots \ge c_{k,{N_{{\rm{a}},k}}}$ and $g_{k,1} \ge \ldots \ge
g_{k,{N_{{\rm{a}},k}}}$. Since ${\bf{A}}_k$ and $\hat{\bf{H}}_{{\rm{eff}},k}[n]\hat{\bf{H}}_{{\rm{eff}},k}^H [n]$ are both positive definite,
$c_{k,i} > g_{k,i}, i=1,\ldots, N_{{\rm{a}},k}$. Calculating the first and second derivation of (\ref{eq16_1}), (\ref{eq13}) is strictly concave
and monotonically increasing. Then Theorem \ref{Theorem1} is verified.

\section*{Acknowledgment}

This work is supported in part by Huawei Technologies Co. Ltd., Shanghai, China, Chinese Important National Science and Technology Specific
Project (2010ZX03002-003) and National Basic Research Program of China (973 Program) 2007CB310602. The authors would like to thank the anonymous
reviewers for their insightful comments and suggestions.


%








\begin{thebibliography}{10}
\bibitem{Fettweis1}
G. Fettweis, Ernesto Zimmermann, "ICT Energy Consumption - Trends
And Challenges", \emph{in proc. of WPMC 2008}

\bibitem{Blume1}
O. Blume, D. Zeller and U. Barth, "Approaches to Energy Efficient
Wireless Access Networks", \emph{in Proceedings of the 4th
International Symposium on Communications, Control and Signal
Processing, ISCCSP 2010}, Limassol, cyprus, 3-5 March 2010



\bibitem{Kim1}
H. Kim, G. de Veciana, "Leveraging Dynamic Spare Capacity in
Wireless System to Conserve Mobile Terminals' Energy,"
\emph{IEEE/ACM Trans. Networking}, vol. 18, no. 3, pp. 802-815, June
2010.

\bibitem{Miao1}
G. W. Miao, N. Himayat, G. Y. Li, and D. Bormann, ¡°Energy-efficient
design in wireless OFDMA,¡± \emph{Proc. IEEE 2008 International
Conference on Communications}, pp. 3307-3312, Beijing , China , May
2008.

\bibitem{Miao2}
G. W. Miao, N. Himayat, G. Y. Li, and A. Swami, ¡°Cross-layer
optimization for energy-efficient wireless communications: a
survey,¡± (invited), \emph{Wiley Journal Wireless Commun. and Mobile
Computing}, vol.9, no.4, pp. 529-542, Apr. 2009.


\bibitem{Miao3}
G. W. Miao, N. Himayat, and G. Y. Li, ¡°Energy-efficient link
adaptation in frequency-selective channels,¡± \emph{IEEE
Transactions on Communications}, vol. 58, no. 2, pp. 545-554, Feb.
2010.


\bibitem{huawei_vtc}
S. Zhang, Y. Chen, and S. Xu, "Improving Energy Efficiency through
Bandwidth, Power, and Adaptive Modulation", \emph{IEEE Proceeding of
2010 Vehicular Technology Conference fall}

\bibitem{Cui}
S. Cui, A. J. Goldsmith, A. Bahai, "Energy-Efficiency of MIMO and
cooperative MIMO Techniques in Sensor Networks," \emph{IEEE Journal
of Selected Areas Commun.}, vol. 22, no. 6, pp. 1089-1098, Aug.
2004.

\bibitem{Kim2}
H. Kim, C.-B. Chae, G. Veciana, and R.W. Heath, "A Cross-Layer
Approach to Energy Efficiency for Adaptive MIMO Systems Exploiting
Spare Capacity," \emph{IEEE Trans. Wireless Commun.}, Vol. 8, No. 8,
August 2009

\bibitem{HSKim}
H. S. Kim, B. Daneshrad, "Energy-Constrained Link Adaptation for
MIMO OFDM Wireless Communication Systems", \emph{IEEE Trans.
Wireless Commun.}, vol.9, no.9, pp.2820-2832, Sep. 2010


\bibitem{Spencer}
Q. H. Spencer, A. L. Swindlehurst, and M. Haardt, ¡°Zero-forcing
methods for downlink spatial multiplexing in multi-user MIMO
channels,¡± \emph{IEEE Trans. Signal Process.}, vol. 52, no. 2, pp.
461-471, Feb. 2004.

\bibitem{Shen}
Z. Shen, R. Chen, J. G. Andrews, R. W. Heath Jr., and B. L. Evans.
"Low complexity user selection algorithms for multiuser MIMO systems
with block diagonalization", \emph{IEEE Trans. Signal Processing},
vol.54, no. 9, pp.3658-3663, Sept. 2006.

\bibitem{Rchen1}
Z. Shen, R. Chen, J. G. Andrews, R. W. Heath, Jr., and B. L. Evans,
"Sum capacity of multiuser MIMO broadcast channels with block
diagonalization," \emph{IEEE Trans. Wireless Commun.}, vol. 6, no.
6, pp. 2040-2045, Jun. 2007.

\bibitem{Rchen2}
R.~Chen, Z.~Shen, J.~G.~Andrews, and R.~W.~Heath~Jr., "Multimode
Transmission for Multiuser MIMO Systems With Block Diagonalization",
\emph{IEEE Trans. Signal Processing}, vol. 56, no. 7, pp. 3294-3302,
July 2008.

\bibitem{Zhang1}
J. Zhang, R. W. Heath Jr., M. Kountouris, and J. G. Andrews, "Mode
Switching for MIMO Broadcast Channel Based on Delay and Channel
Quantization," \emph{EURASIP Journal on Advances in Signal
Processing}, Volume 2009, Article ID 802548, 15 pages

\bibitem{Zhang2}
J. Zhang, J. G. Andrews, and R. W. Heath Jr., "Block Diagonalization
in the MIMO Broadcast Channel with Delayed CSIT", \emph{in IEEE
proc. of Globecom 2009}, pp. 1-6, Nov. 2009

\bibitem{Zhang3}
J. Zhang, M. Kountouris, J. G. Andrews and R. W. Heath Jr.,
"Multi-mode Transmission for the MIMO Broadcast Channel with
Imperfect Channel State Information," avaiable online at
http://arxiv.org/abs/0903.5108v2

\bibitem{Xu2}
J. Xu and L. Qiu, "Robust Multimode Selection in the Downlink Multiuser MIMO Channels with Delayed CSIT", in \emph{ IEEE proc. of ICC 2011}, pp.
1-5, June 2011


\bibitem{Arnold}
O. Arnold, F. Richter, G. Fettweis and O. Blume, "Power Consumption
Modeling of Different Base Station Types in Heterogeneous Cellular
Networks", \emph{in Proceedings of the ICT MobileSummit (ICT
Summit'10), Florence, Italy, 16. - 18. June 2010}



\bibitem{Telatar}
I. E. Telatar, ¡°Capacity of multi-antenna Gaussian channels,¡±
\emph{Europ. Trans. Telecommun.}, vol. 10, pp. 585-595, Nov. 1999.


\bibitem{Yoo}
T.Yoo and A.J.Goldsmith, ''Capacity and power Allocation for fading
MIMO channels with channel estimation error'', \emph{IEEE Trans.
Inf. Theory}, vol.52, no.5, pp.2203-2214, May 2006

\bibitem{Rapajic}
P. Rapajic and D. Popescu, ¡°Information capacity of a random
signature multiple-input multiple-output channel,¡± \emph{IEEE
Trans. Commun.}, vol. 48, pp. 1245-1248, Aug. 2000.

\bibitem{Jie_Xu}
J. Xu, L. Qiu and C. Yu, "Link Adaptation and Mode Switching for the Energy Efficient Multiuser MIMO Systems", submitted to \emph{IEICE Trans.
Commun.}, [available online], {\url{http://home.ustc.edu.cn/~suming/}}

\end{thebibliography}
\end{document}